\documentclass[pra,aps,twocolumn,amsmath,amssymb,superscriptaddress]{revtex4}

\usepackage{amstext}
\usepackage{graphicx}
\usepackage{color}
\usepackage{epstopdf}

\begin{document}

\title{Entanglement of an Impurity in a Few-Body One-Dimensional Ideal Bose System}

\author{M. A. Garc\'{\i}a-March} 
\affiliation{ICFO-Institut de Ci\`encies Fot\`oniques, The Barcelona Institute of Science and Technology, Av. C.F. Gauss 3, 08860 Castelldefels (Barcelona), Spain}
\author{A. S. Dehkharghani} 
\affiliation{Department of Physics and Astronomy, Aarhus University, DK-8000 Aarhus C, Denmark}
\author{N. T. Zinner} 
\affiliation{Department of Physics and Astronomy, Aarhus University, DK-8000 Aarhus C, Denmark}

\begin{abstract}
We study the correlation between an impurity and a small ensemble of bosonic particles in one dimension. Our study analyzes the one-body density matrix and calculates the corresponding von Neumann entanglement entropy as a function of the interaction strength between the impurity and the bosons when all particles have the same mass. We show that the entropy grows very fast for small and moderate interaction strength and then increases slowly toward the strongly interacting regime. Then we study the effect over the quantum correlations of a mass imbalance between the impurity and the bosons. In the strongly interacting case, we discover that when the impurity is much heavier than the bosons, then we have the least possible correlation. However, the entropy tops its maximum when the mass ratio is between 3 and 4 in the case where there are four bosonic particles and then falls off to its minimum for higher mass imbalance.
\end{abstract}

\maketitle
\section{Introduction}

The interest for one dimensional ultra cold quantum systems has grown considerably in the last few years. This is partly due to the realization of such systems in highly controllable environments using cold atomic 
gases~\cite{paredes2004,kino2004,kinoshita2006,haller2009,serwane2011,gerhard2012,wenz2013} and partly due to the revelation of interesting quantum phenomena that are unique to one dimension (1D). Some of the fundamental 1D effects include spin-charge separation, which has been studied recently for both fermions and bosons~\cite{recati2003}. The spin separation is usually caused by the well-known Pauli principle for fermions, however in Bose mixtures, where the bosonic system consists of two-component subsystems, the separation has also been found. In that sense, the bosonic systems are more interesting than fermionic ones because here, in addition to inter-species interaction, one can also tune the intra-species interaction, which is the interaction between the same type of particles. The latter is clearly not an option for fermions due to the Pauli principle (when considering only $s$-wave interactions). In Bose mixtures it has been shown that a Ferromagnetic ground state occurs when the intra- and inter-species interactions are identical \cite{eisenberg2002,nachtergaele2005,deuretzbacher2008,massignan2015,yang2015}.

A particular case of Bose mixtures where one single atom of one kind, usually called the impurity, interacting with a number of quantum particles of other kind, also called majority particles, is also an interesting field in 1D systems. Just to mention a few studies and effects in this area one can mention the Landau-Pekar polaron \cite{landau1933,pekar1948} and a magnetic impurity in a metal resulting in the Kondo effect \cite{kondo1964}. Moreover, the impurity atom interacting with a Bose-Einstein condensate has been found to have a bound ground state in which the impurity is self-localized in the so-called polaron-like state \cite{cucchietti2006, kalas2006, bruderer2008,Tempere2013}. There has been considerable recent interest in the physics of polarons in 
quasi-condensates \cite{rath2013,li2014,grusdt2015}, also in 
the case where the mass of the impurity is different from the mass of the bosons \cite{mehta2014,artem2015b,mehta2015}.
In a recent study the crossover between few- and many-body behaviors in Bose mixtures in the equal mass case was explored \cite{dehkharghani2015a}. Moreover, a new method has been developed to study the general case of the many body Bose polarons \cite{dehkharghani2015b}. Studying such systems and their behavior can help to understand the physics behind some general phenomena in polaron and many-body physics.

Here we extend the studies done before by investigating how correlated the impurity and the majority particles are with each other. The correlation is investigated in two cases; one as a function of interaction strength between the impurity and majority particles and the other as a function of mass ratio between the two components.

The paper is organized as follows. In Sec.~\ref{sec:Hamil} we introduce the model we study and explore the dynamics of a two species system consisting of one atom of one type and up to 4 atoms of other type. In this section we focus on the same mass case but vary the interaction strength between the two species. In Sec.~\ref{sec:Mass} we extend our discussion to the mass imbalanced case and highlight the main effects by showing the behavior of the one-body density matrix for the impurity and majority particles. In addition we calculate the entanglement between them and show how the highest occupied orbitals evolve as we vary the interaction strength and mass ratio.

\section{Hamiltonian of the system and mass-balanced wave functions }
\label{sec:Hamil}
We consider a one-dimensional mixture of $N_\mathrm{A}$ identical bosons of one kind, A, with coordinates $x_i$, $i=1,\dots,N_\mathrm{A}$, and one atom
of kind B, with coordinate $y$. We assume that the $N_\mathrm{A}$ identical bosons do not interact among themselves. We consider contact repulsive interactions between the $N_\textrm{A}$ bosons of type A and the additional B atom, which 
we call {\it the impurity}. We model the interactions by a delta function of strength given by the coupling constant $g$. 

We assume the same trapping oscillator frequency $\omega$ for the bosons and the impurity, although we note that 
the formalism we use can in principle treat also different oscillator frequencies. 

We assume that the mass of the bosons, $m_{\mathrm{A}}$ can be different from that of the impurity, $m_{\mathrm{B}}$. In this situation, the Hamiltonian reads
\begin{align}
\label{eq:Hamiltonian}
H &= \frac{1}{m_{\mathrm{BA}}}\left(-\frac{1}{2}\frac{ d^2}{dy^2}\right) + m_\mathrm{BA} \frac{1}{2} y^2 \\
&+ \sum_{i=1}^{N_\mathrm{A}} \left[-\frac{1}{2}\frac{d^2}{dx_i^2} + \frac{1}{2} x_i^2\right] + g \sum_{i=1}^{N_\mathrm{A}} \delta(x_i-y),\nonumber
\end{align}
where $m_{\mathrm{BA}}\equiv m_{\mathrm{B}}/m_{\mathrm{A}}$. Here, we scaled all energies by $\hbar\omega$ and all distances by the harmonic oscillator length $a_{\mathrm {ho}}=\sqrt{\hbar/m_{\mathrm{A}}\omega}$. Thus, the coupling constant $g$ is scaled by $\hbar\omega a_{\mathrm {ho}}$. 

As a benchmark, let us first briefly discuss the case in which $m_{\mathrm{BA}}=1$ and $g$ is tuned from zero to infinity. 
In the non-interacting case, $g=0$, the ground state is non-degenerate, and its wave function is real, positive, without zeros, and symmetric under the exchange of all atoms. It is given by
\begin{equation}
\Psi_{\rm g.s.}(x_1,\dots,x_{N_{\mathrm{A}}},y)\! =\! \psi_0(x_1) \dots\psi_0(x_{N_{\mathrm{A}}})\psi_0(y) \,,
\end{equation}
with Gaussian $ \psi_0(x)=\pi^{-1/4}\exp(-x^2/2)$, and energy $E_{\rm g.s.}= (N_{\mathrm{A}}+1)/2$. 
When $g\to\infty$, the wave function has to vanish at all points where $x_i=y$, $\forall i$. A wave function as an ansatz fulfilling this condition could be \cite{Zollner:08a,Garcia-March:14,Garcia-March:14b} 
\begin{align}
\label{eq:analyticalwf_3}
\Psi_{\rm g.s.}^{2,\rm bos}(x_1,\dots,x_{N_{\mathrm{A}}},y) 
&\propto \exp{[ - \frac {1}{2} (x_1^2+\dots+x_{N_\mathrm{A}}^2+y^2)]} \\
&\times |x_1-y|\;\cdots |x_{N_\mathrm{A}}-y|.\nonumber
\end{align}
This wave function is zero whenever one of the bosons and the impurity are in the same position. This ansatz is real and positive, that is, it has zeros but not changes of sign. The fact that the impurity is distinguishable from the rest of bosons means that there is no symmetrization condition under the interchange of any of the bosons and the impurity. The modulus in Eq.~(\ref{eq:analyticalwf_3}) enforces that when interchanging any A atom with the impurity, the wave function is even. 
But one can have an odd wave function under the interchange of the impurity and any A atom. These kind of functions have been discussed for $N_{\mathrm{A}}=2 $ in~\cite{Zinner:13, Garcia-March:14b} 
(see also the analysis based in symmetry arguments given in~\cite{harshman2012,Harshman:15,yurovsky2014}).
It was shown previously \cite{Zinner:13,dehkharghani2015a} that the two-fold degeneracy comes from the fact that the impurity in the equal mass case is pushed to the side of the trap. It may sit on either the left- or right-hand side of all the bosons. However, since we need to conserve parity the two degenerate states will be even and odd linear combinations of having it localized on the left or the right \cite{dehkharghani2015a}.

\begin{figure}
\includegraphics[width=.75\columnwidth]{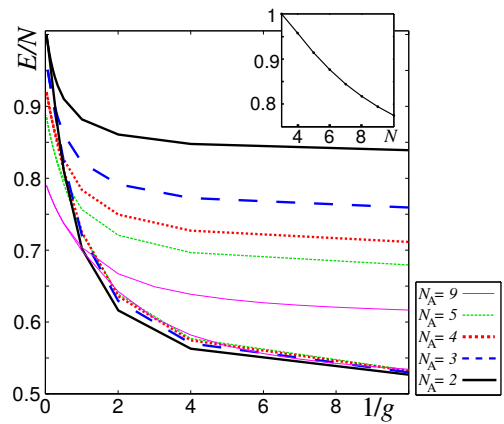}
\caption{Energy per total number of atoms of the ground and first excited state as a function of the inverse of the coupling constant, for $N_{\mathrm{A}}=2,3,4,5 $ and 9. A quasi-degenerate pair occurs in the lowest part of the spectrum as $1/g\to 0$. Inset: energy per atom for $g\to\infty$ as a function of the total number of atoms, $N$.}
\label{fig1}
\end{figure}

In Fig.~\ref{fig1} we show how this quasi-degenerate energy pair appears as $g$ is increased, for $N_{\mathrm{A}}=2,3,4 $ and 9. We calculated this spectra with a many-mode direct diagonalization algorithm as the one described in~\cite{Garcia-march:12,Garcia-March:13} for $N_{\mathrm{A}}=2,3,4$. We also calculated for larger number of atoms with the method described in~Appendix~\ref{sec:app1} and~\cite{dehkharghani2015b}, obtaining the same quasi-degenerated pair in the lower part of the spectra for larger number of bosons. In Fig.~\ref{fig1} we show the results for $N_{\mathrm{A}}=9$. Note that for small $g$, the energy per atom of the ground state is $1/2$ while the one of the excited state is $(N_{\mathrm{A}}/2+3/2)/(N_{\mathrm{A}}+1)$. For large $g$, the energy of the quasi-degenerate pair is smaller as $N_{\mathrm{A}}$ is increased, as shown in the inset of Fig.~\ref{fig1} (see \cite{dehkharghani2015a}). 

\begin{figure}
\vspace{1cm}
\includegraphics[width=\columnwidth]{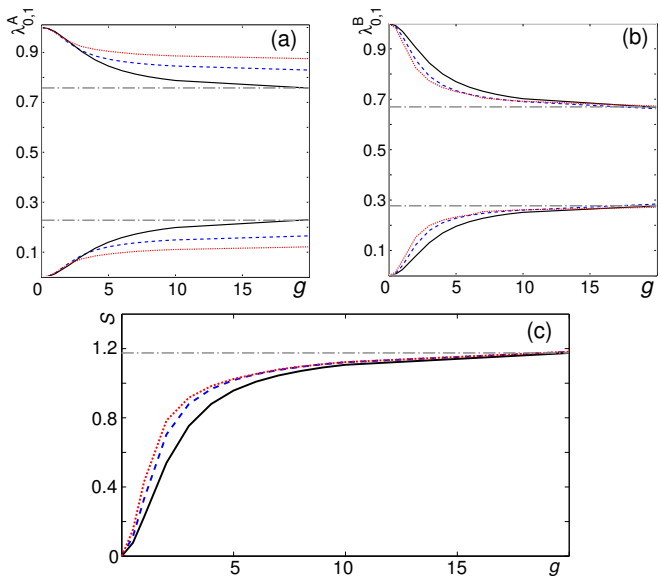}
\caption{(a) First and second largest natural orbit occupations for the bosons and (b) for the impurity, for $N_{\mathrm{A}}=2,3$ and $4$ (solid black, dashed blue and dotted red, respectively). (c) Entanglement entropy for the same cases. The dashed-dotted lines are the values taken from the analytical results for $1/g=0$ and $N_\mathrm{A}=2$.
\label{fig2}}
\end{figure}

In this work we are interested in the complicated correlations and coherences that are present in the system due to the interactions between the bosons and the impurity. To this end it is natural to compute the one-body density matrices (OBDM) for an A atom and for the impurity one. These read
\begin{equation}
\rho^{{\mathrm{A}}}(x,x')\!=\!N_{\mathrm{A}}\!\!\int\!\! d x_2\dots d x_{N_{\mathrm{A}}}d\,y\Psi(x,\dots)\Psi(x',\dots), 
\end{equation}
and
\begin{equation}
\rho^{{\mathrm{B}}}(y,y')\!=\!\int\!\! d x_1\dots d x_{N_{\mathrm{A}}}\Psi(x_1,\dots,y)\Psi(x_1,\dots,y'). 
\end{equation}
This bosonic system is not in an ideal Bose system state (product of ground state single-particle wave functions) when $g$ becomes sufficiently large. This conclusion can be reached from diagonalization of the OBDM, which produces the natural orbitals and their occupations, $\lambda_i$. The largest occupation of a natural orbital for the bosons, $\lambda_0^{\mathrm{A}}$, is reduced as $g$ is increased [see Fig.~\ref{fig2}(a)]. As one increases the number of bosons, the system is less disturbed by the presence of the impurity, and the value of $\lambda_0$ for large $g$ gets closer to one. On the other hand, the largest occupation of a natural orbit for the impurity, $\lambda_0^{\mathrm{B}}$, is also reduced as $g$ is increased [see Fig.~\ref{fig2}(b)]. Based on our investigation for $N_\mathrm{A}\leq 4$ the final value of $\lambda_0^{\mathrm{B}}$ does not seem to depend on the number of atoms in A. 
Moreover, if we consider two different parts of the system, one being the bosons and the other the impurity, one can see that there are strong quantum correlations built up among the two as $g$ is increased. In order to show this effect, we calculate the entanglement entropy, which is defined as $S(\rho^{{\mathrm{B}}})=-\mbox{Tr}[\rho^{{\mathrm{B}}}~\mbox{log}_2~\rho^{{\mathrm{B}}}]$, which can be obtained from the natural orbits occupation as
$ S(\rho^{{\mathrm{B}}})=-\sum_i \lambda_i \mbox{log}_2 \lambda_i\;$. Note that this formula provides the entanglement entropy between the bosons and the impurity because it corresponds to tracing out all degrees of freedom associated with A. As shown in Fig.~\ref{fig2}(c) this entropy grows as $g$ is increased, showing that indeed the interactions are responsible for the built up of correlations between the bosons and the impurity. Also we observe that the entanglement entropy grows steeply for $g<5$. Finally, this growth becomes sharper for larger values of $N_{\mathrm{A}}$. 

\begin{figure}
\begin{tabular}{cc}
\hspace{0.5cm}\includegraphics[width=0.45\columnwidth]{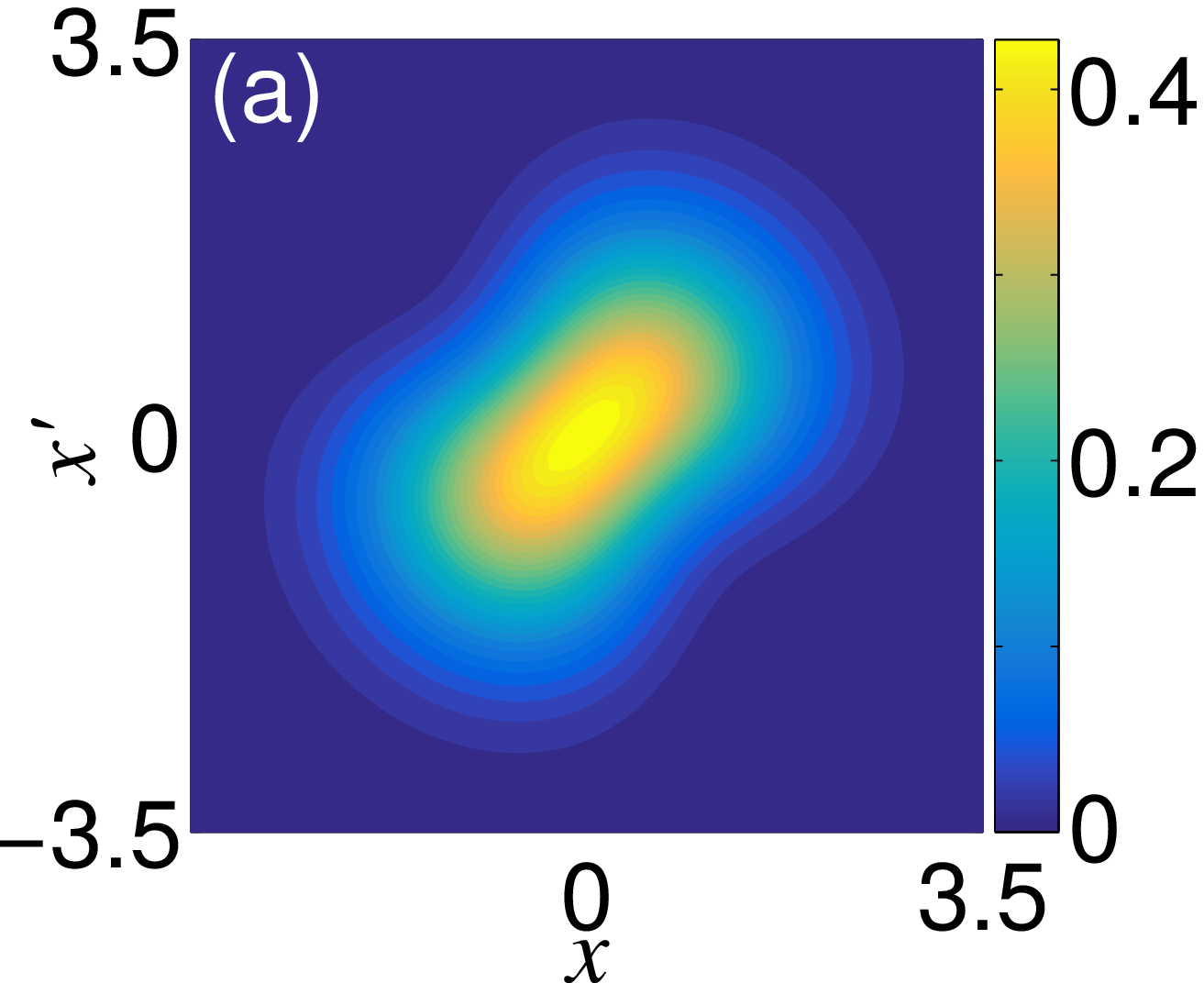}&\hspace{-0.25cm}\includegraphics[width=0.45\columnwidth]{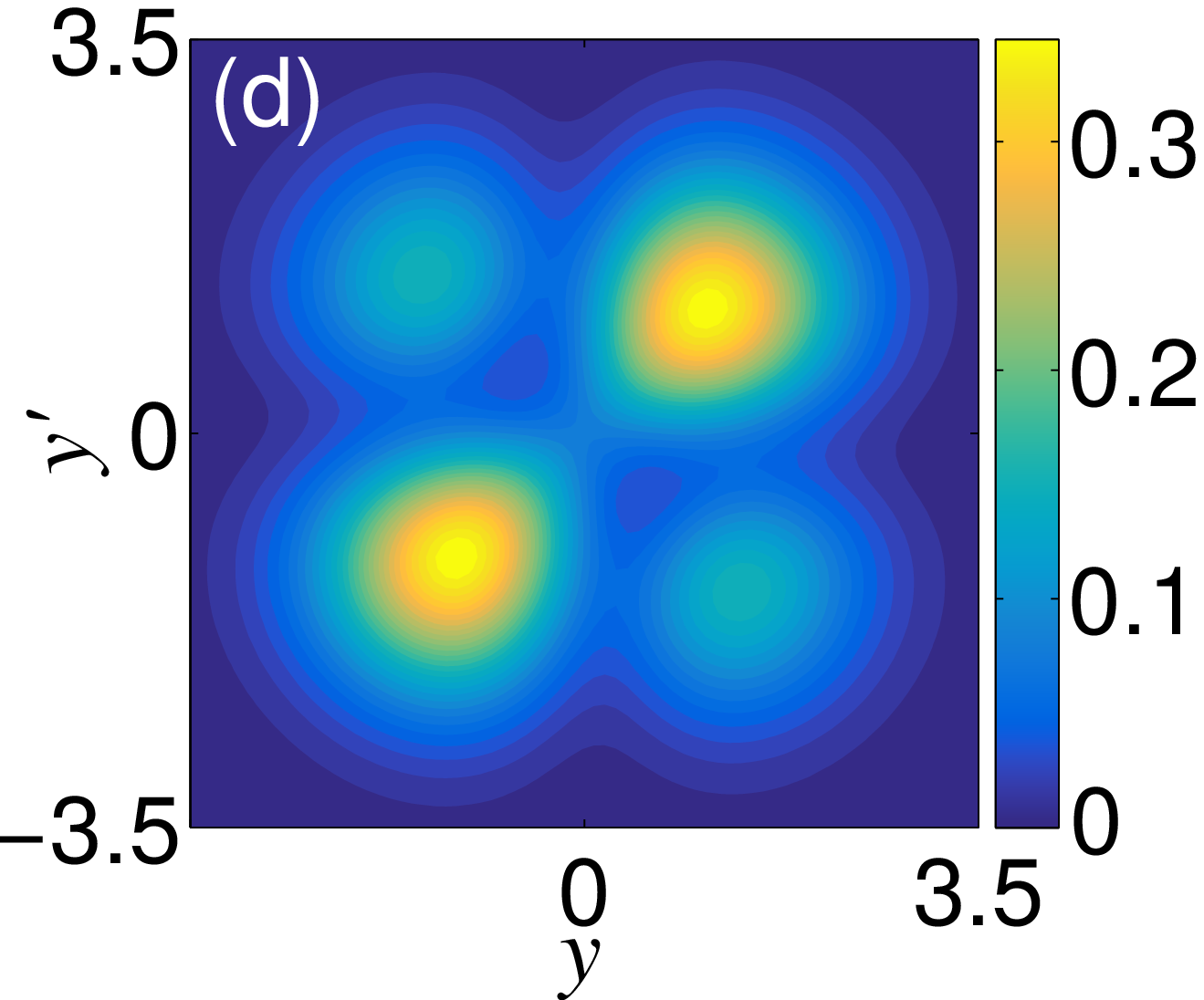}\\
\hspace{0.5cm}\includegraphics[width=0.45\columnwidth]{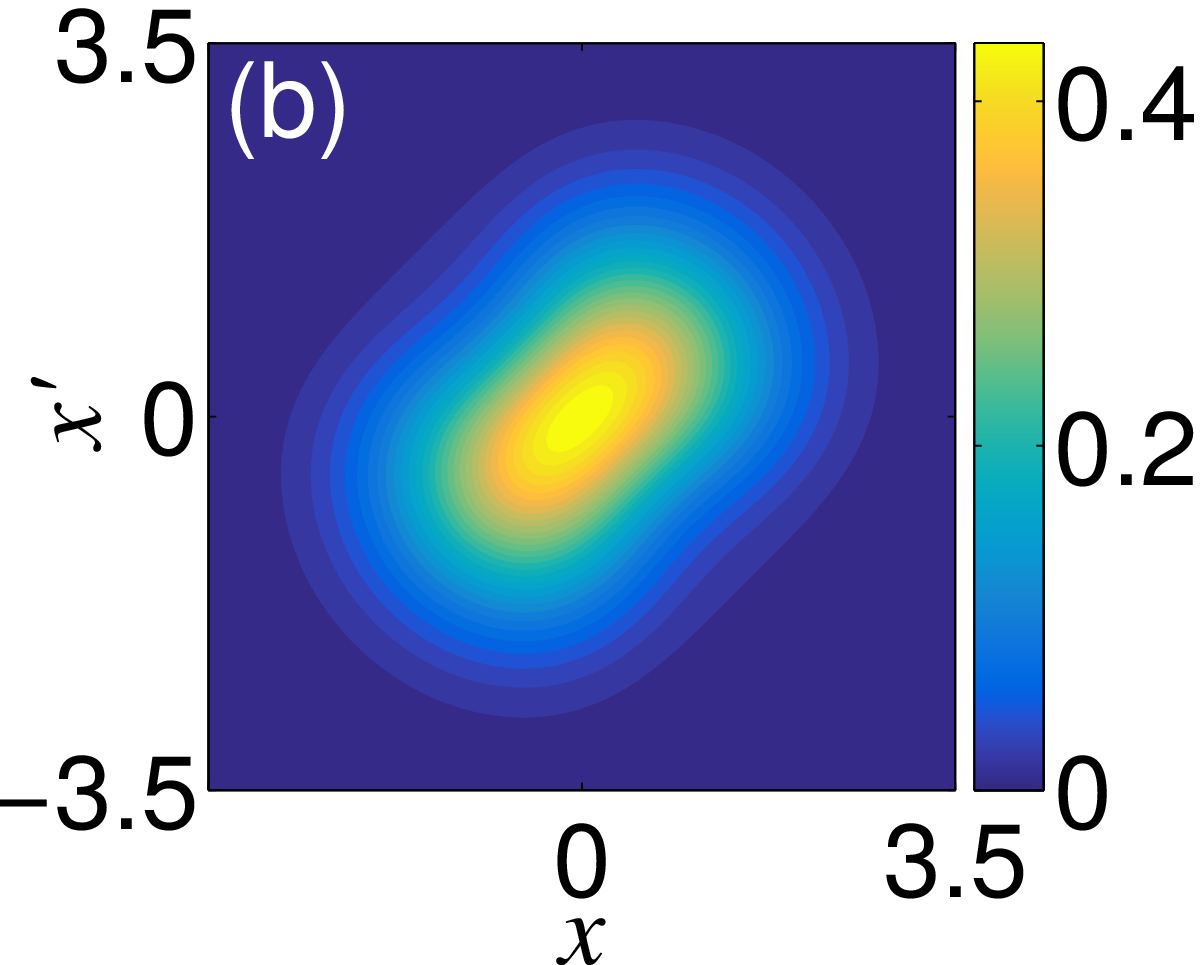}&\hspace{-0.25cm}\includegraphics[width=0.45\columnwidth]{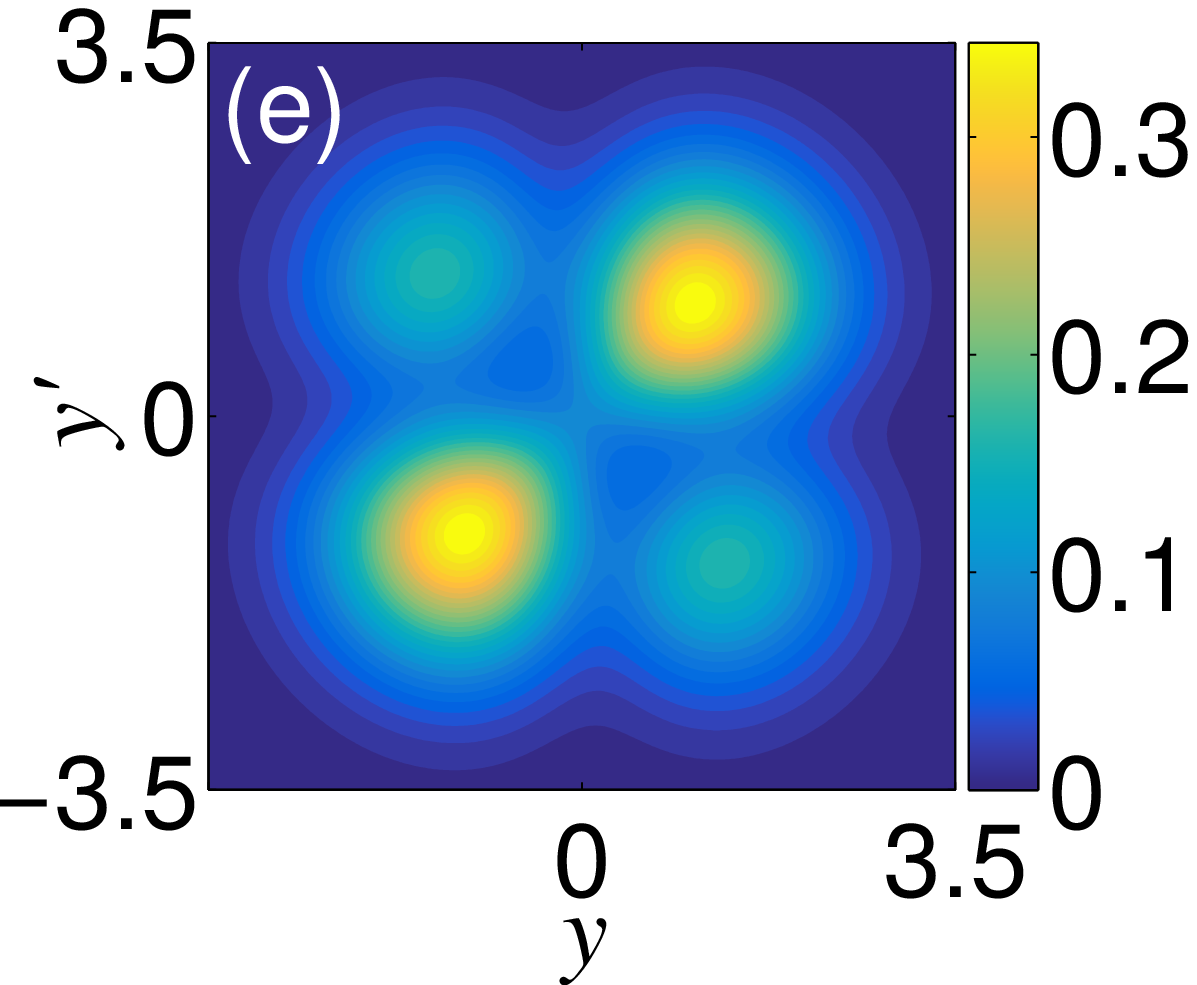}\\
\hspace{0.5cm}\includegraphics[width=0.45\columnwidth]{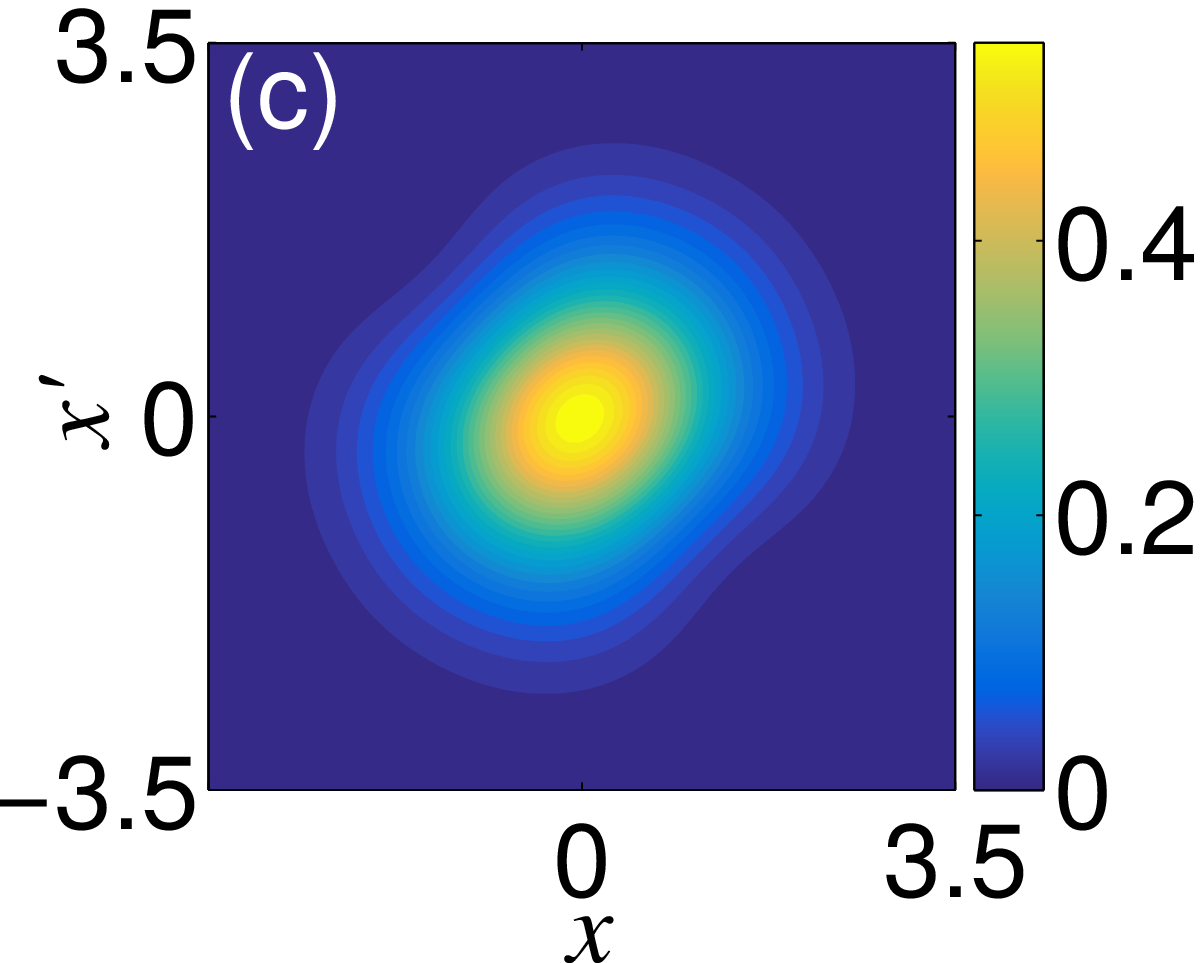}&\hspace{-0.25cm}\includegraphics[width=0.45\columnwidth]{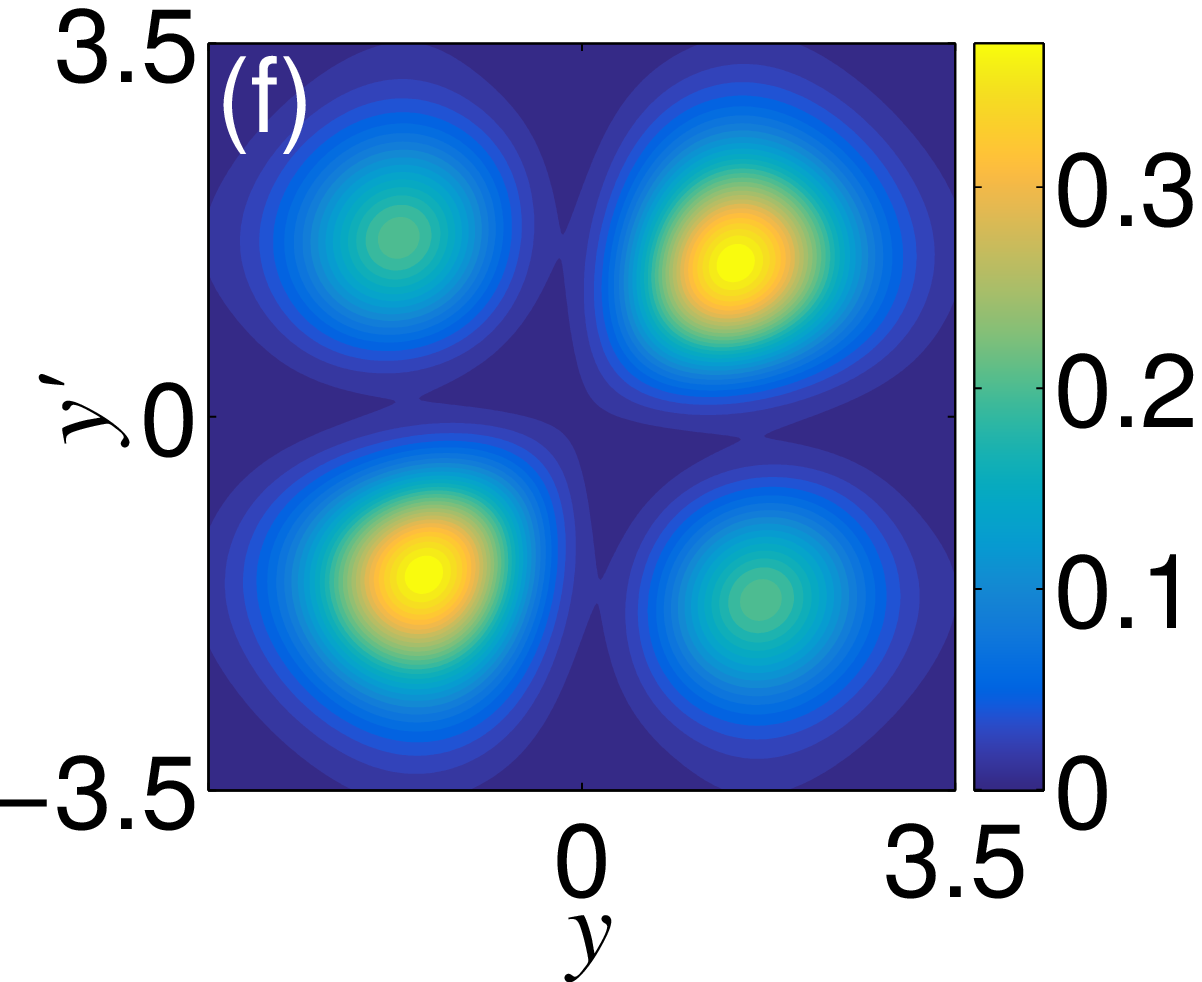}
\end{tabular}
\caption{One body density matrices for the bosons (left column) and the impurity (right column) for the ground state. a) and d) show the analytical result with infinite interaction for $N_{\mathrm{A}}=2$, while b), c), e) and f) show the numerical results for $N_{\mathrm{A}}=2$ and $4$ when $g=20$.
\label{fig3}}
\end{figure}

In Fig.~\ref{fig3} a) and d) we provide the analytical OBDM for the $N_{\mathrm{A}}=2$ case with infinite interaction \cite{harshman2012,Zinner:13}. In Fig.~\ref{fig3} c), d), e) and f) we show the numerically calculated OBDMs of the ground state for the bosons with $N_{\mathrm{A}}=2, 4$ and for the impurity. The numerical OBDMs are represented for $g=20$, which is a value of the coupling constant large enough to have strong correlations between both species (see Fig.~\ref{fig2}). The diagonal of these matrices, that is when $x=x'$ or $y=y'$, is the corresponding density profile. 
As shown, for large $g$, the density profile of the impurity corresponds to it being located in the left or in the right side of the bosons. 
For two bosons the OBDM of the bosons gets more distorted when compared to a Gaussian profile than for four bosons, and indeed gets more Gaussian for $N_{\mathrm{A}}=4$. This shows that the effect of the impurity on the bosons cloud is smaller as A gets more populated. In turn, for larger $N_{\mathrm{A}}$ the two peaks in the density profile for the impurity are displaced further away from the center of the trap. For smaller $g$ the overlap of the density profiles of both species is larger but also the entanglement entropy is smaller, thus showing less correlations between the impurity and the bosons.

\section{Correlations and entanglement entropy for mass imbalanced systems}
\label{sec:Mass}

As $g$ is increased, the entanglement entropy shown in Fig.~\ref{fig2}(c) between the impurity and majority atoms grows, showing that strong correlations between these two parts are built up. 
When $g$ is large enough to have large correlations between both species, one may expect that, if one allows for the impurity to be heavier, there should be some value of the mass of the impurity in which it is heavy enough as to localize in the center of the trap. 
This would effectively imply a mass-driven transition between edge localization and central localization of the impurity. A similar transition driven by mass-imbalance has been seen in two-component fermionic systems \cite{pecak2015}.
Then, in the limit of large ratios $m_{\mathrm{BA}}$, the entanglement entropy may drop down and density separation may occur between the impurity and the bosons. Let us note here that increasing the trapping frequency for B will also force the impurity to locate in the centre of the trap. We expect that this alternative strategy gives similar qualitative results. 
In the following, we discuss how the correlations, coherences and densities of the bosons and the impurity behave when one considers increased mass ratios, $m_{\mathrm{BA}}$. 

\begin{figure}
\begin{tabular}{cc}
\includegraphics[width=0.45\columnwidth]{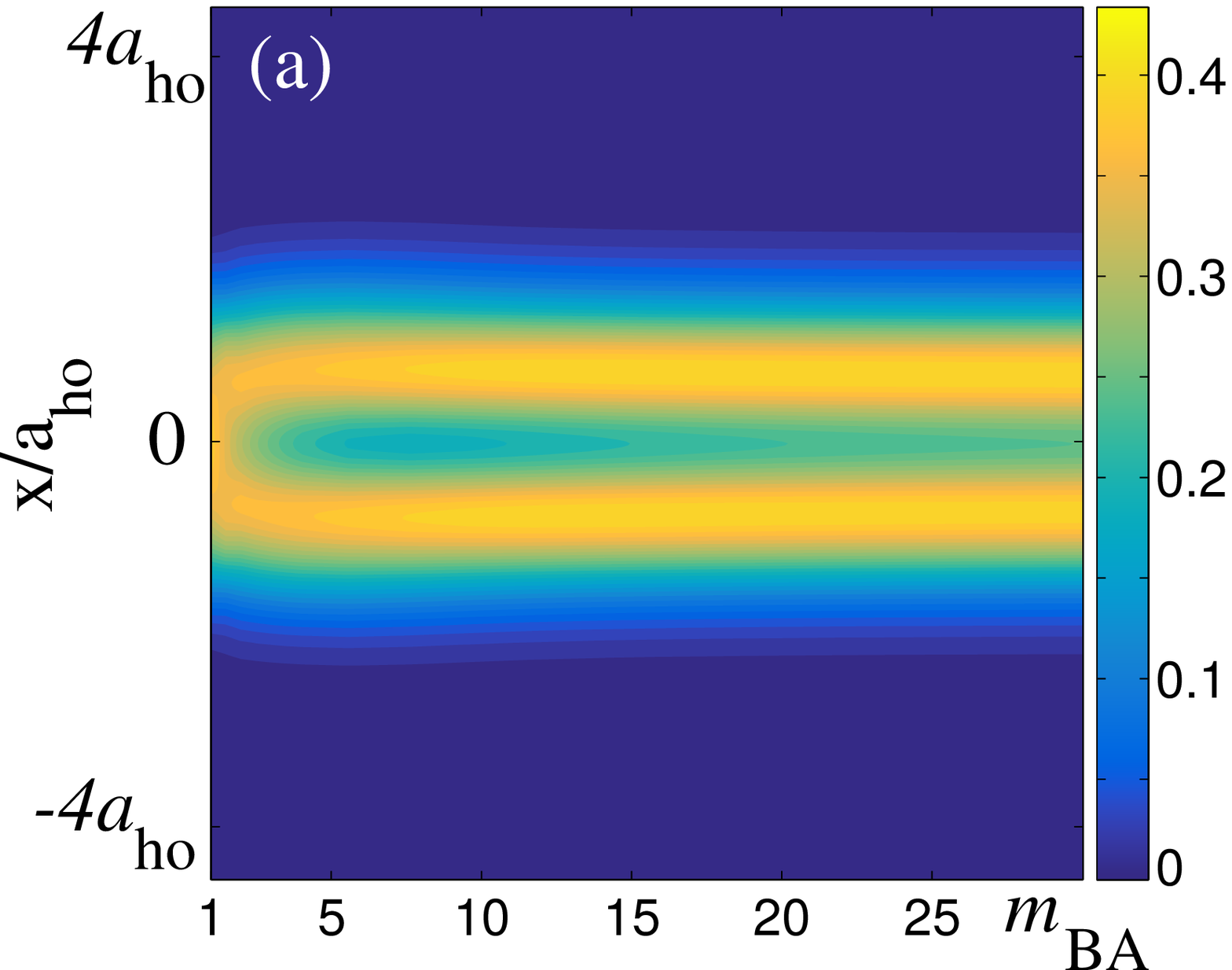}&
\includegraphics[width=0.43\columnwidth]{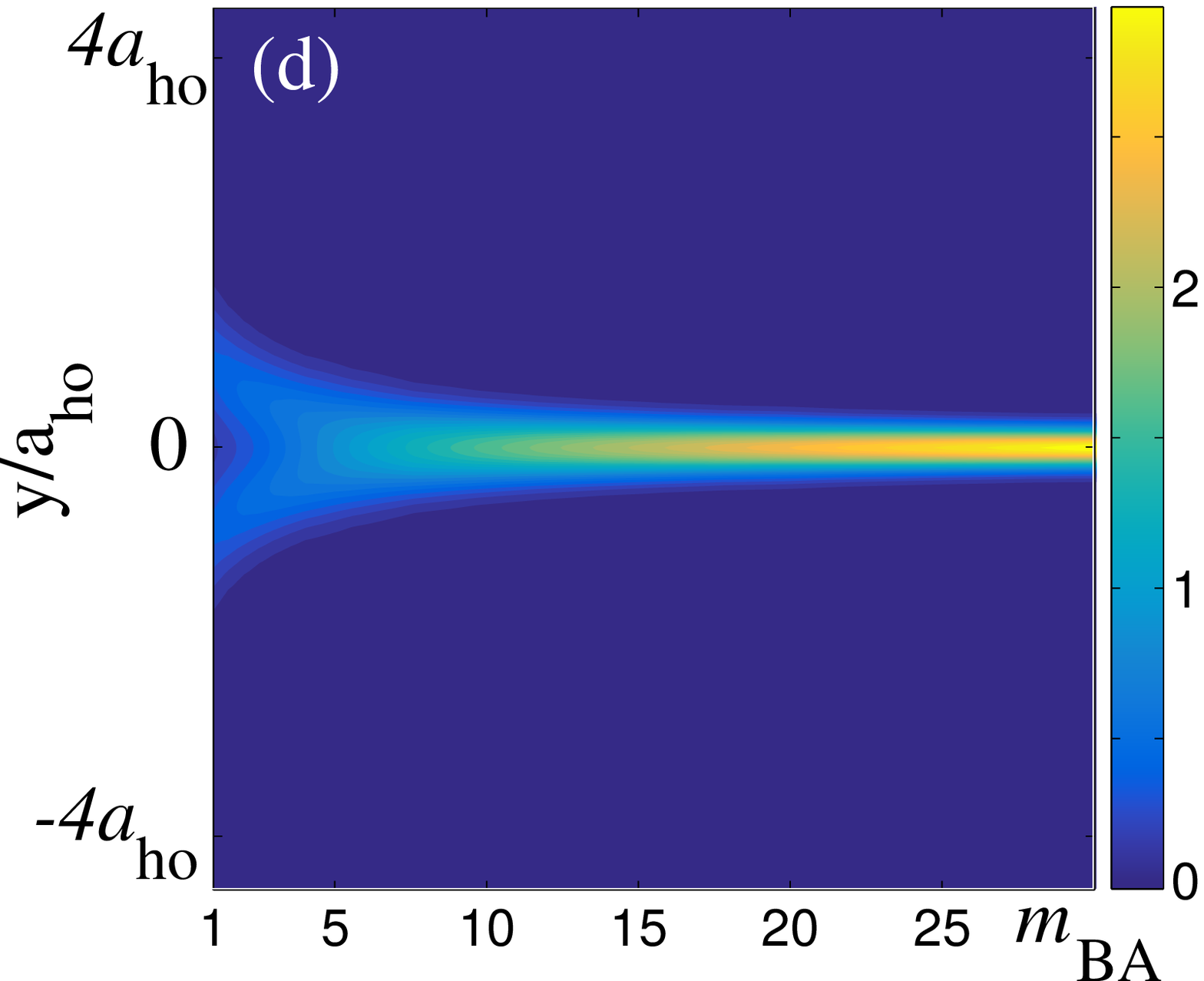}\\
\includegraphics[width=0.45\columnwidth]{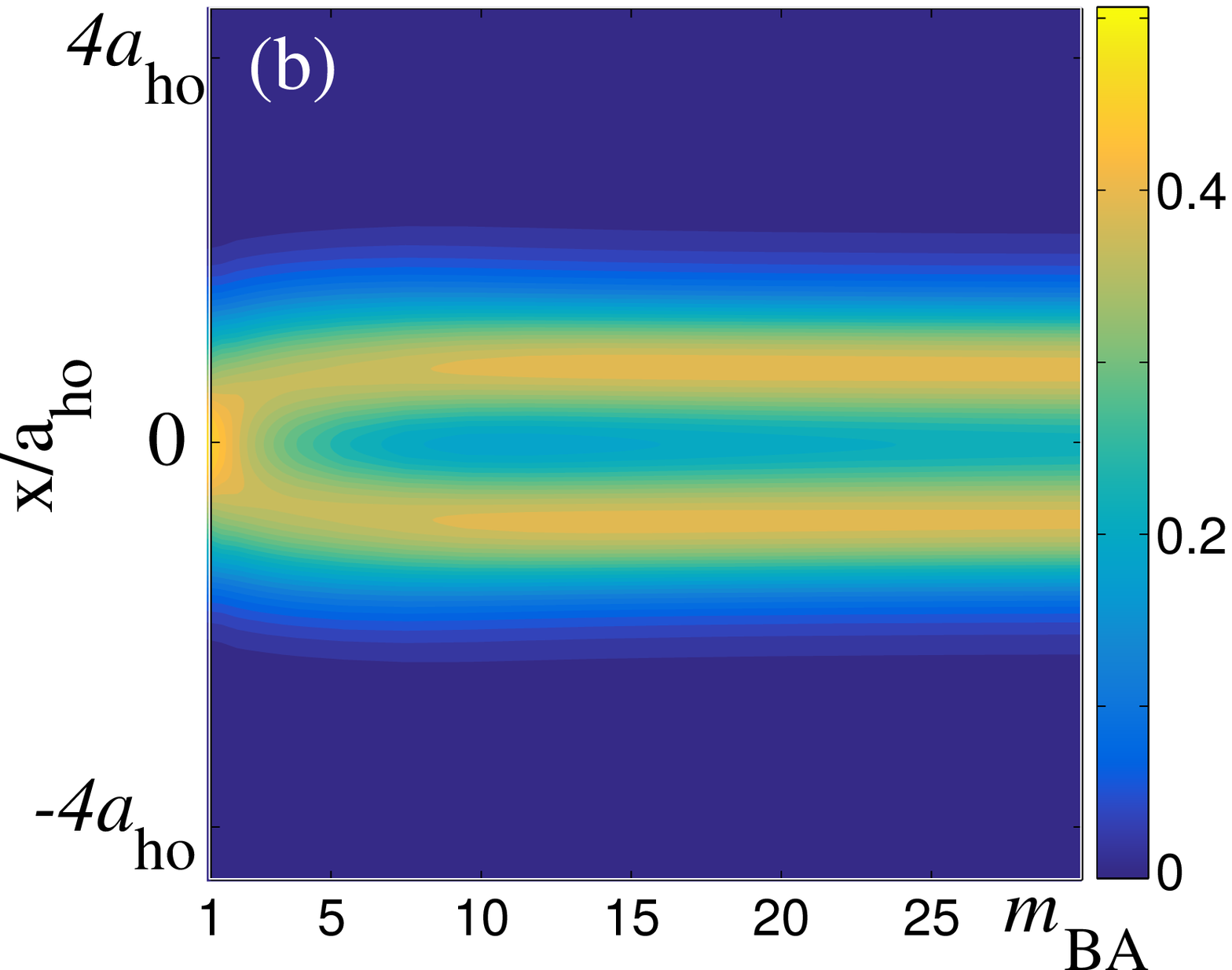}&
\includegraphics[width=0.43\columnwidth]{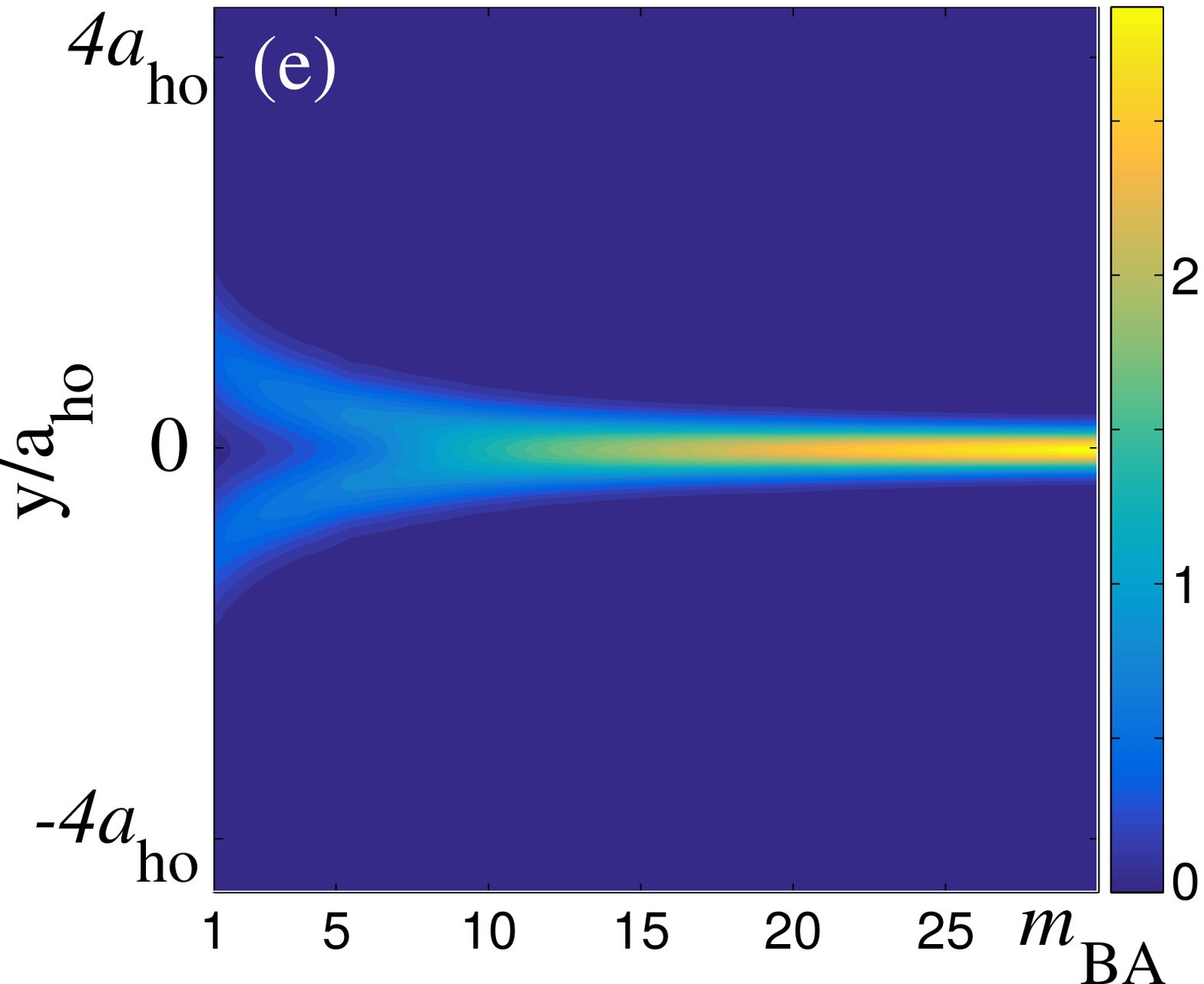}\\
\includegraphics[width=0.45\columnwidth]{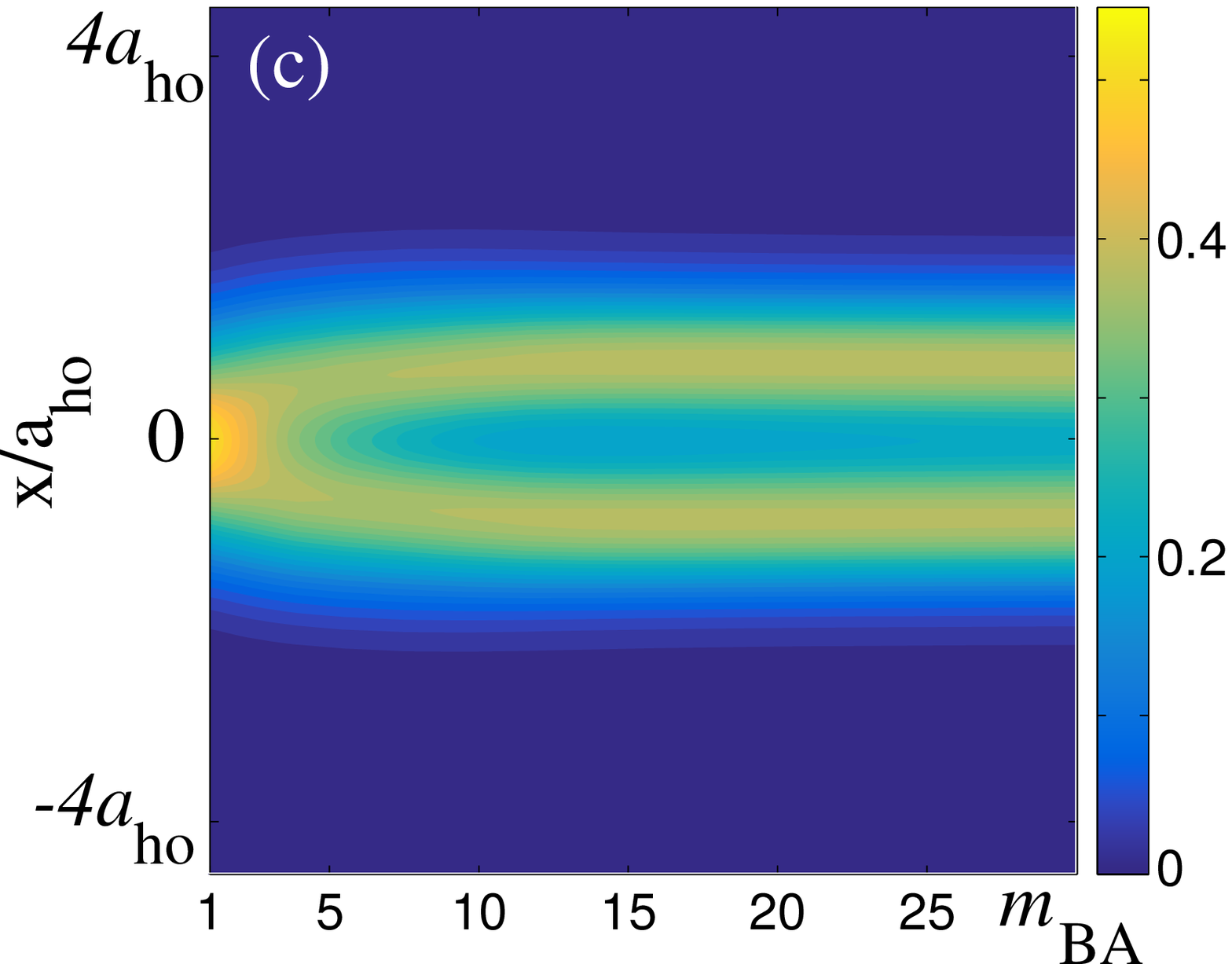}&
\includegraphics[width=0.43\columnwidth]{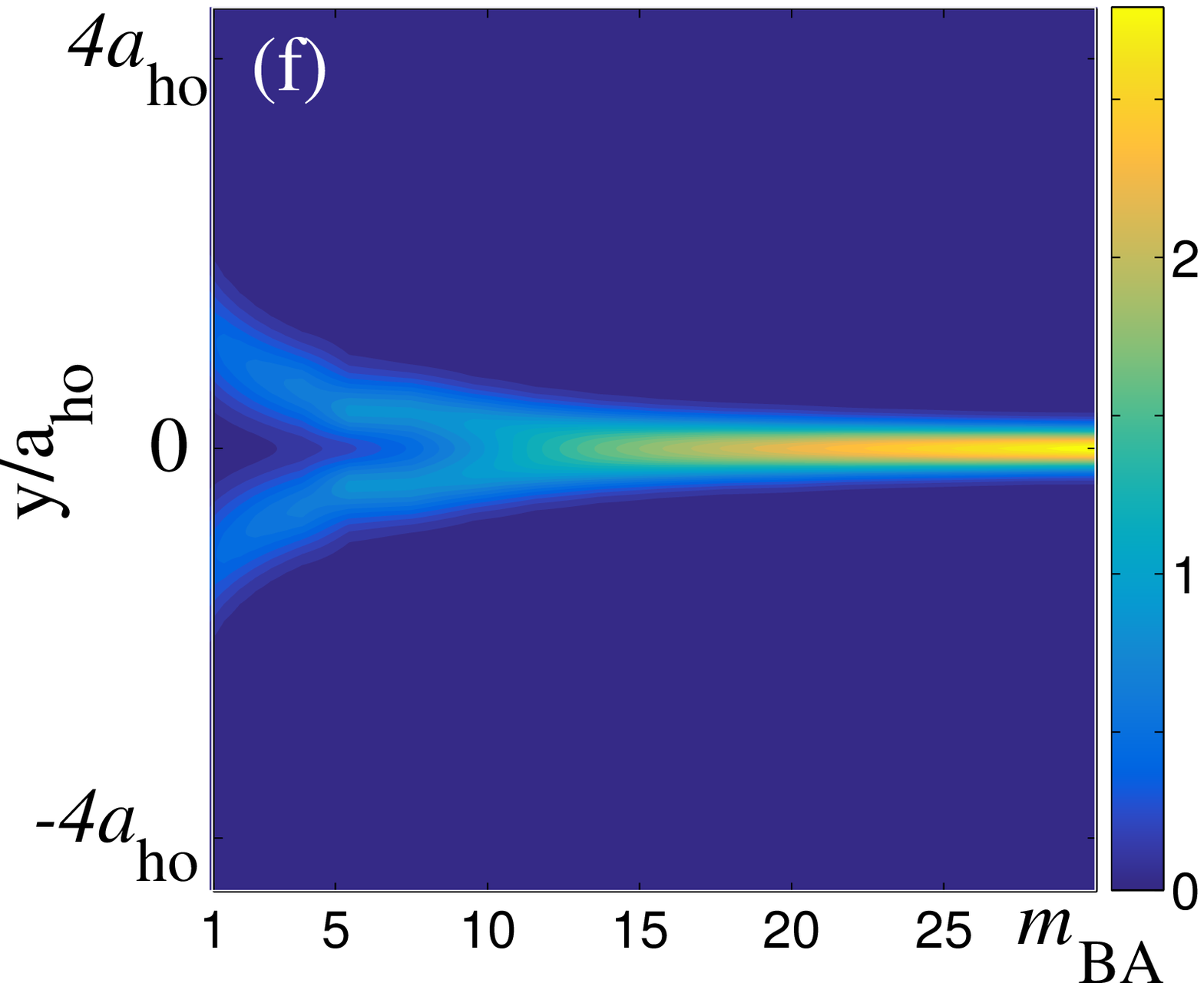}
\end{tabular}
\caption{Density profile of the bosons (left column) and the impurity (right column) as a function of the mass ratio, for $g=10$ and different number of atoms [(a) and (d) for $N_{\mathrm{A}}=2$, (b) and (e) for $N_{\mathrm{A}}=3$, (c) and (f) for $N_{\mathrm{A}}=4$]. Distances are in units of the corresponding oscillator length, $a_{\mathrm {ho}}$, which in the case of the impurity changes as $m_\mathrm{B}$ is increased. As shown, as the mass ratio $m_\mathrm{BA}$ is increased, the impurity tends to localize in the center of the trap while the A atomic cloud shows at dip in the center of the trap.}
\label{fig4}
\end{figure}

The first effect of considering a $m_{\mathrm{BA}}$ slightly greater than one is to break the quasi-degeneracy present in the limit of large $g$. Then, for $m_{\mathrm{BA}}>1$ the ground state is no longer quasi degenerate, even for large $g$. In Fig.~\ref{fig4} we show how the density profile for the impurity and for the bosons changes as the impurity is assumed to be heavier for a large value of $g$ and different values of $N_{\mathrm{A}}$. As expected the impurity tends to localize in the center of the trap for large values of $m_{\mathrm{BA}}$ and the density profiles for the bosons show a minimum coinciding with the center of the trap, where the impurity localizes. This  minimum is smaller for larger values of $N_{\mathrm{A}}$. 
In a previous work \cite{loft2015}, it was shown that in a three-body system with two identical fermions and an impurity the transition of the impurity from the edge to the center of the trap takes place for any infinitesimally small mass-imbalance when $g\to\infty$. In the present case with identical non-interacting bosons we observe that the transition happens at some finite $m_{\mathrm{BA}}>1$ for large but finite coupling constant (the results shown in Fig.~\ref{fig4} have large $g=10$ but not infinite $g$). The fact that the bosons are non-interacting is a clear difference to the fermionic case~\cite{Giraud2009}. Here one needs to consider an intricate competition 
of kinetic, interactions and trap energy. In particular, in order for the impurity to move to the center, it has to push the two bosons out
to the edge, thus increasing their kinetic and trap energy. This may require a finite mass difference to be favorable. Thus, we should not be surprised to see that this transition happens gradually for non-interacting bosons, and occurs for larger mass
imbalance when there are more bosons that have to get out of the way to make room for the impurity in the center of the trap.
The numerical results for $g=10$ provided here are in any case not sufficient to conclude that the transition occurs for 
any $m_{\mathrm{BA}}>1$ at $g\to\infty$. This will be an interesting topic for future work.

\begin{figure}
\includegraphics[width=0.99\columnwidth]{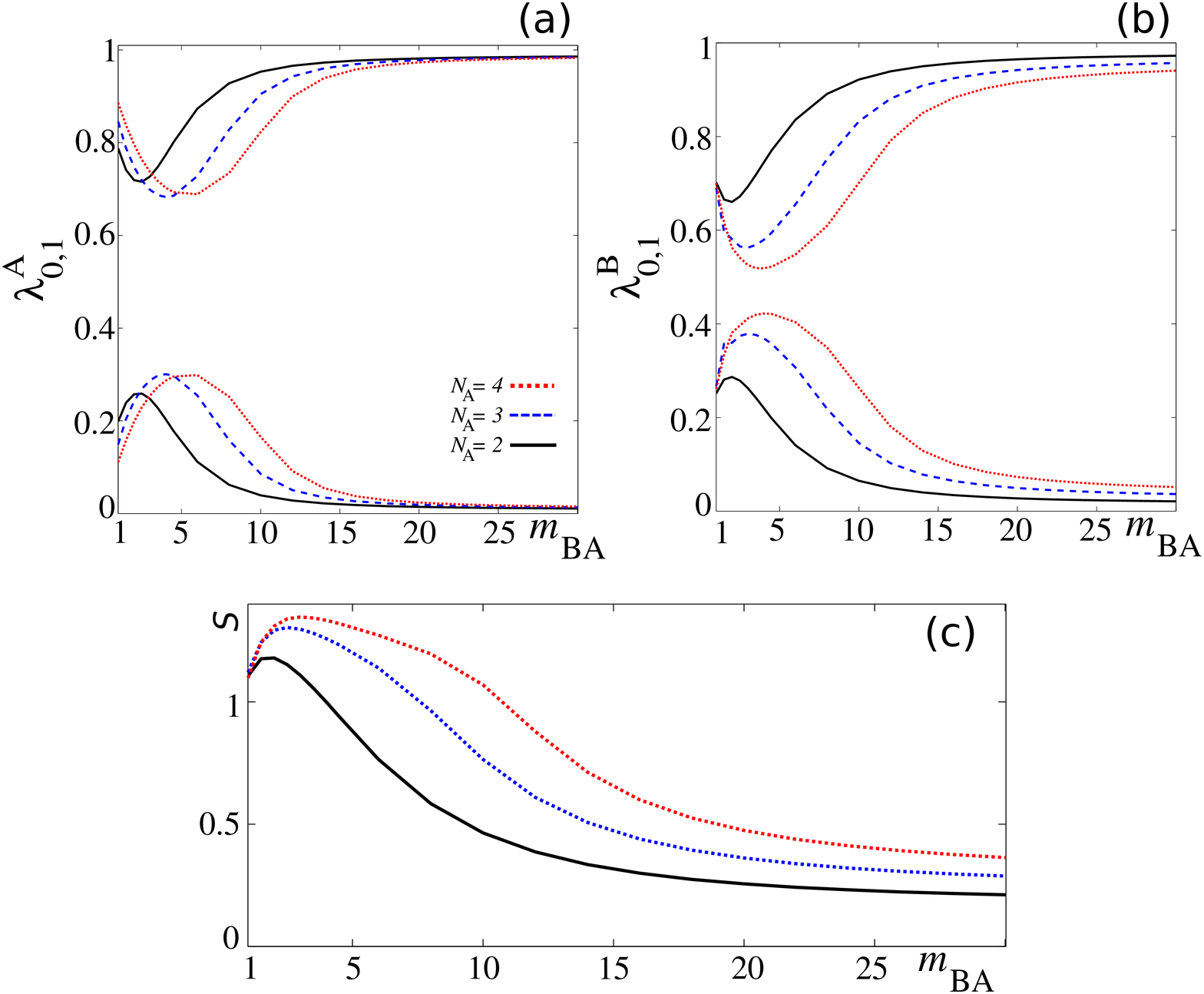}
\caption{Largest and second largest natural orbits occupations for the bosons (a) and the impurity (b), as a function of the mass ratio, for $g=10$ and different $N_{\mathrm{A}}$. (c) Entanglement entropy as a function of the mass ratio. The localization of the impurity in the center of the trap is accompanied with an increase in the natural orbital occupations and a reduction in the entanglement entropy.}
\label{fig5}
\end{figure}

\begin{figure}
\begin{tabular}{cc}
\hspace{0.5cm}
\includegraphics[width=0.45\columnwidth]{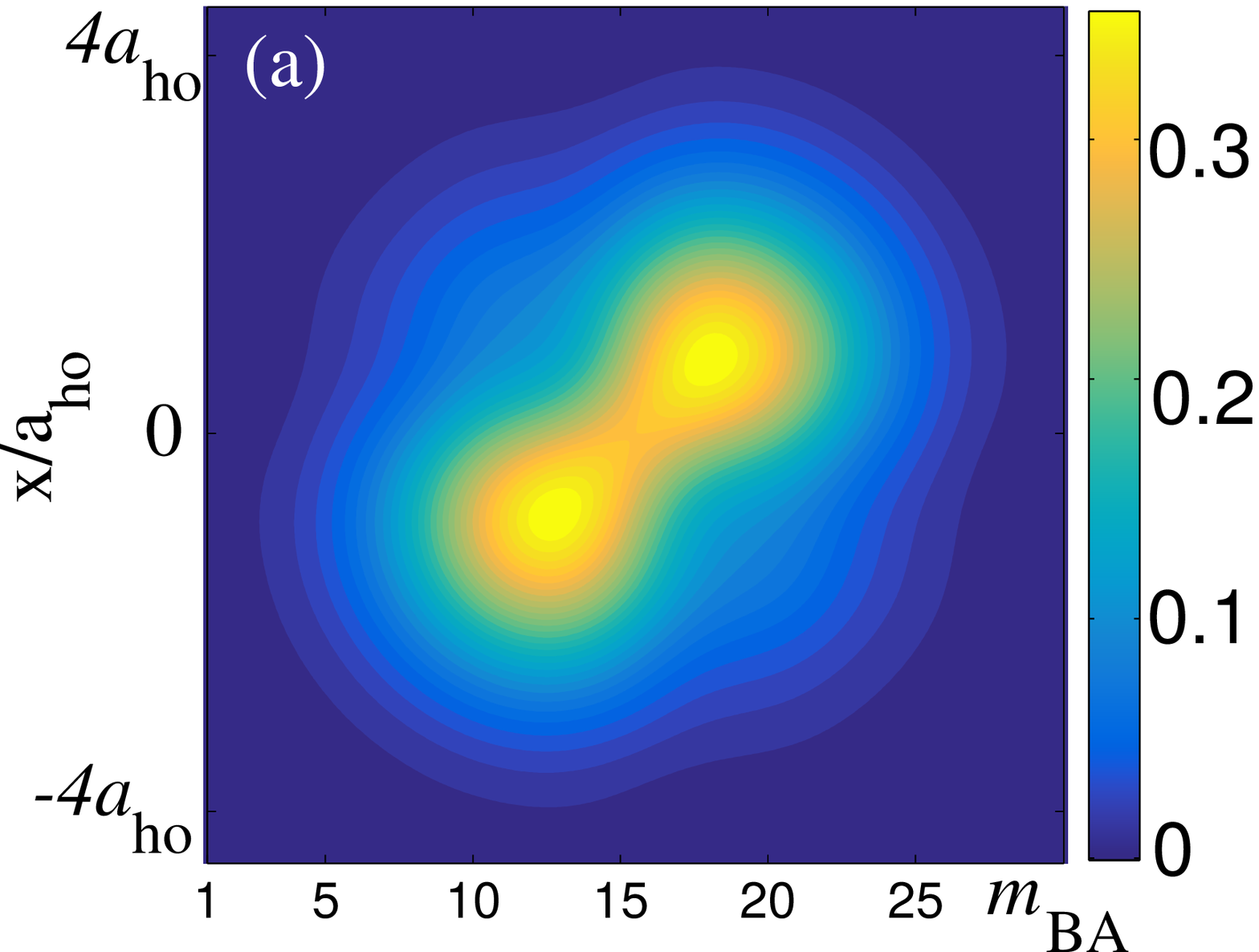}&
\hspace{-0.25cm}
\includegraphics[width=0.45\columnwidth]{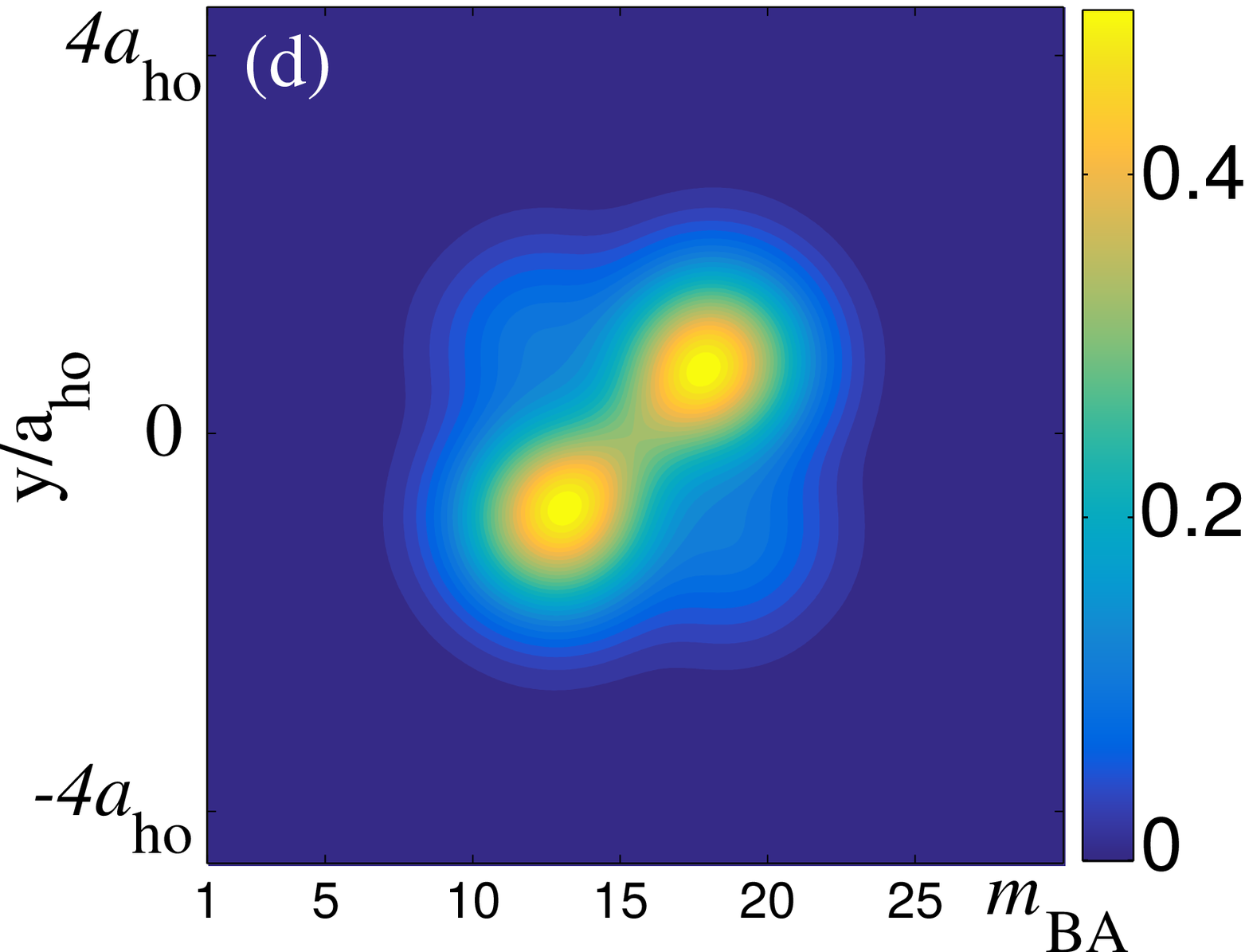}\\
\hspace{0.5cm}
\includegraphics[width=0.45\columnwidth]{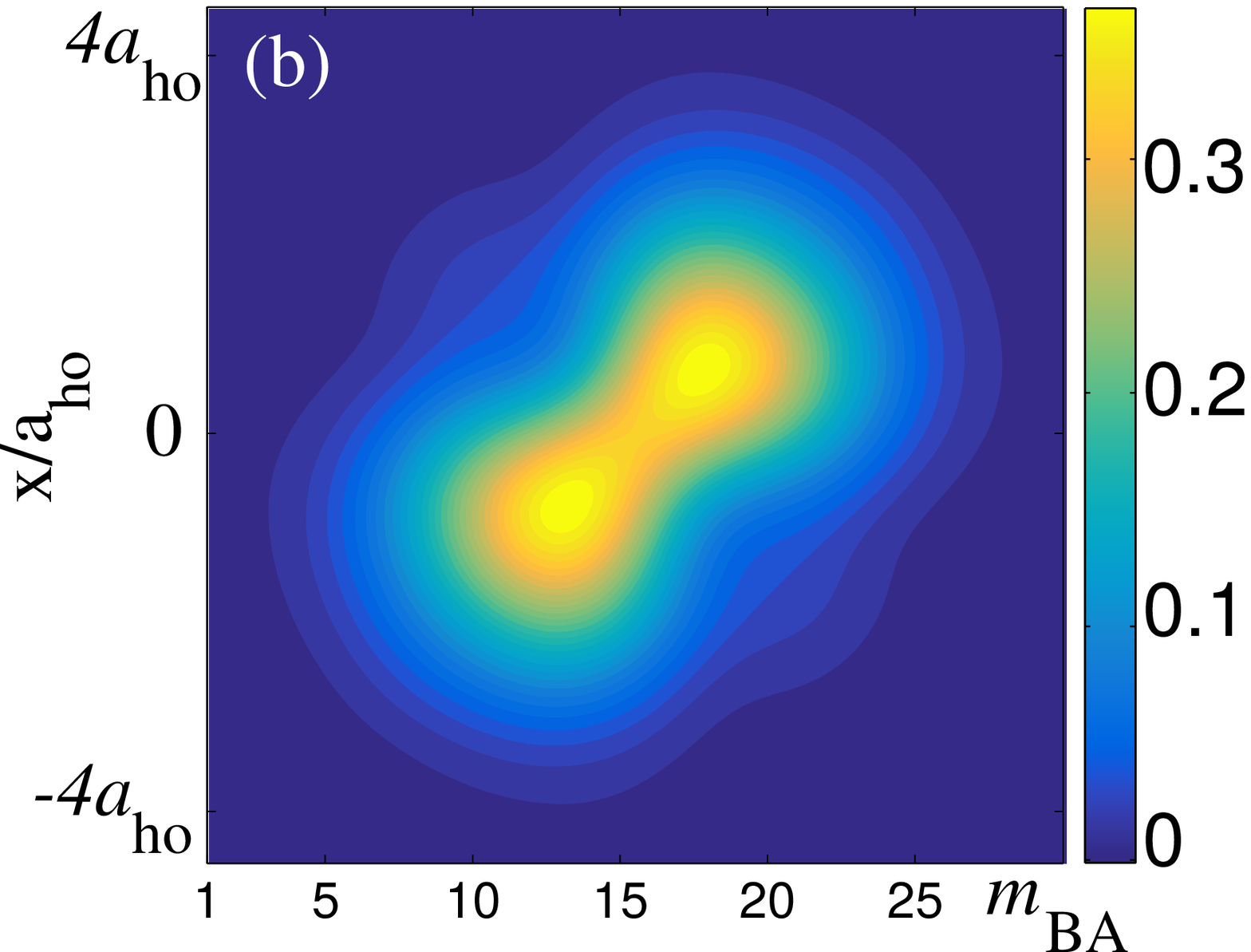}&
\hspace{-0.25cm}
\includegraphics[width=0.45\columnwidth]{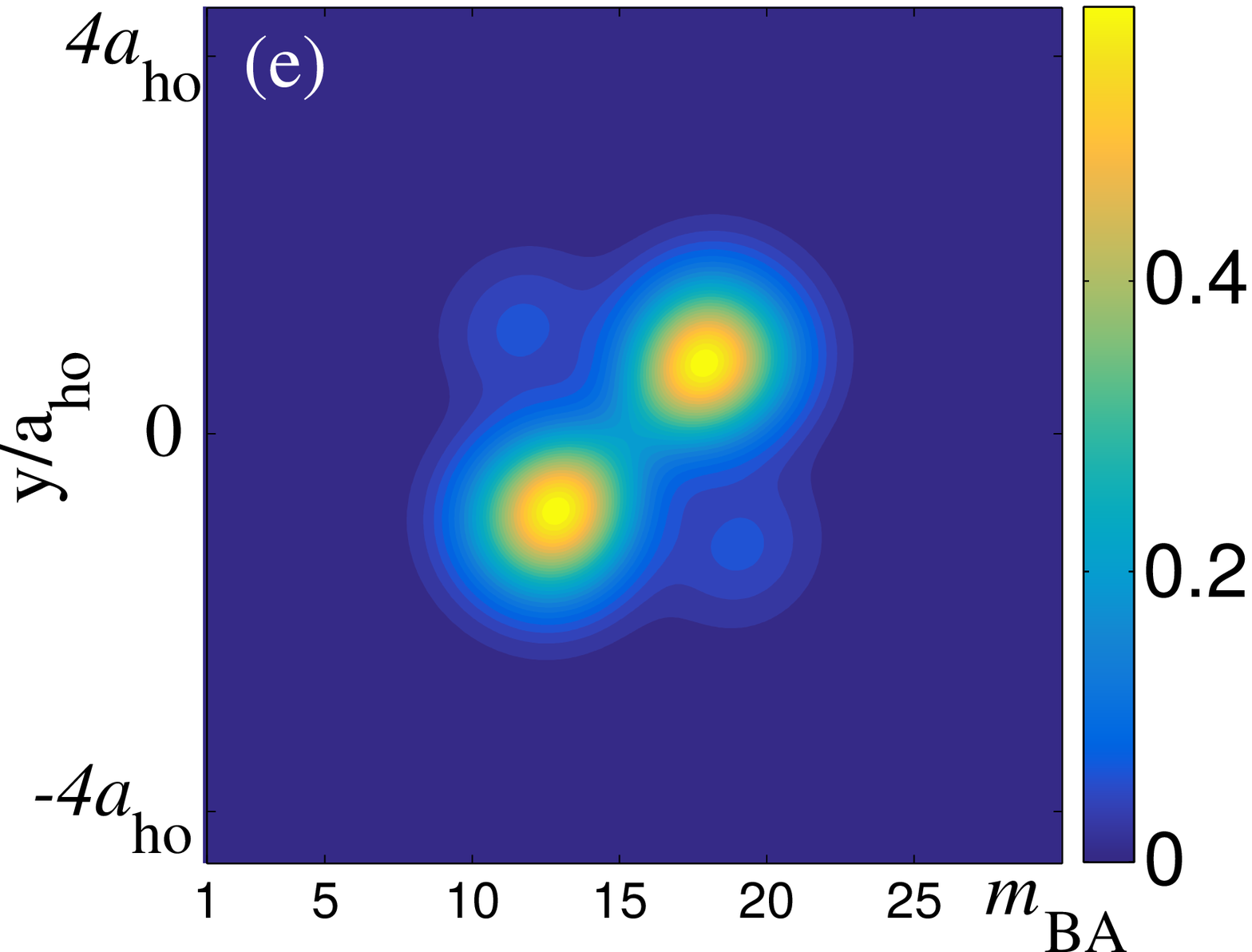}\\
\hspace{0.5cm}
\includegraphics[width=0.45\columnwidth]{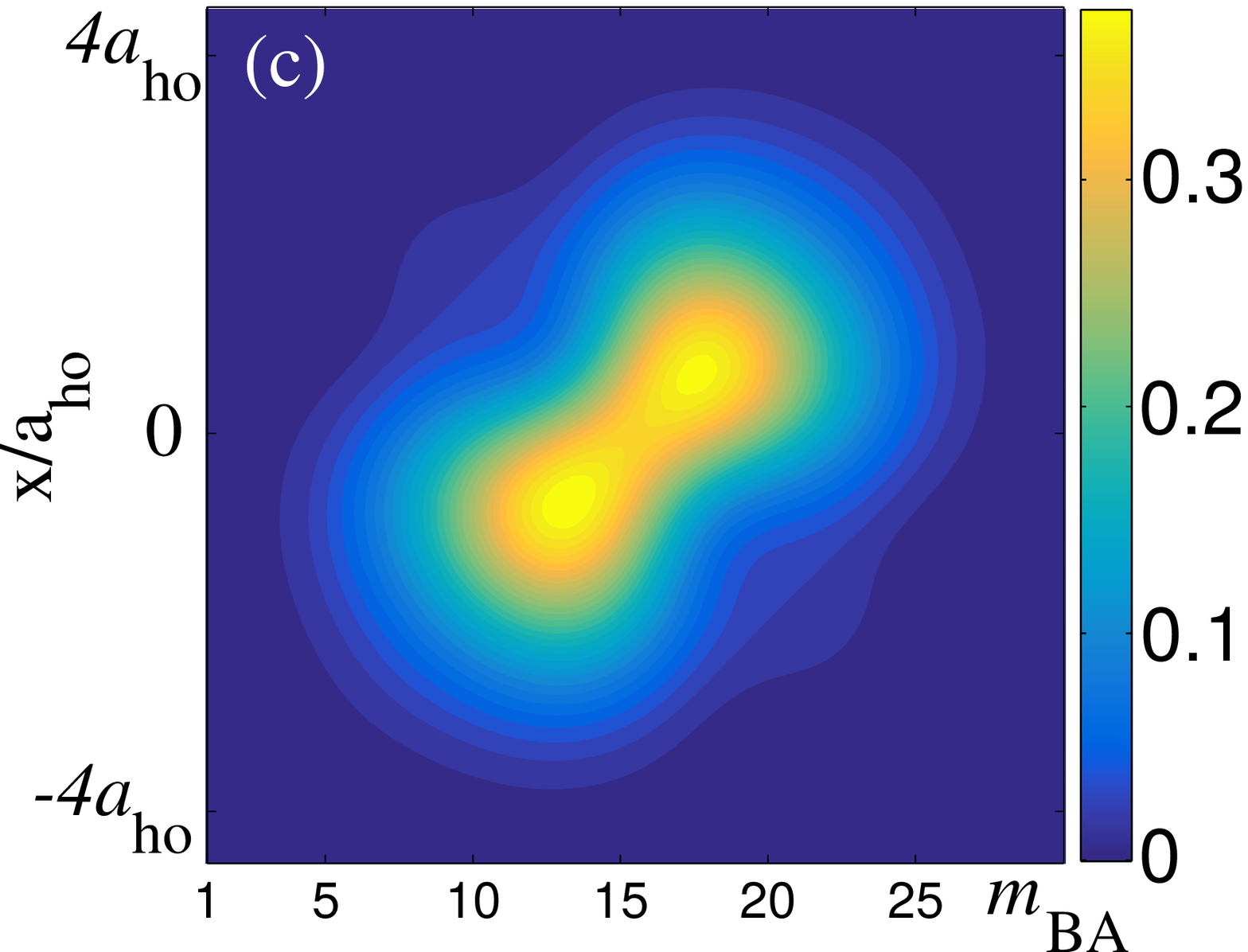}&
\hspace{-0.25cm}
\includegraphics[width=0.45\columnwidth]{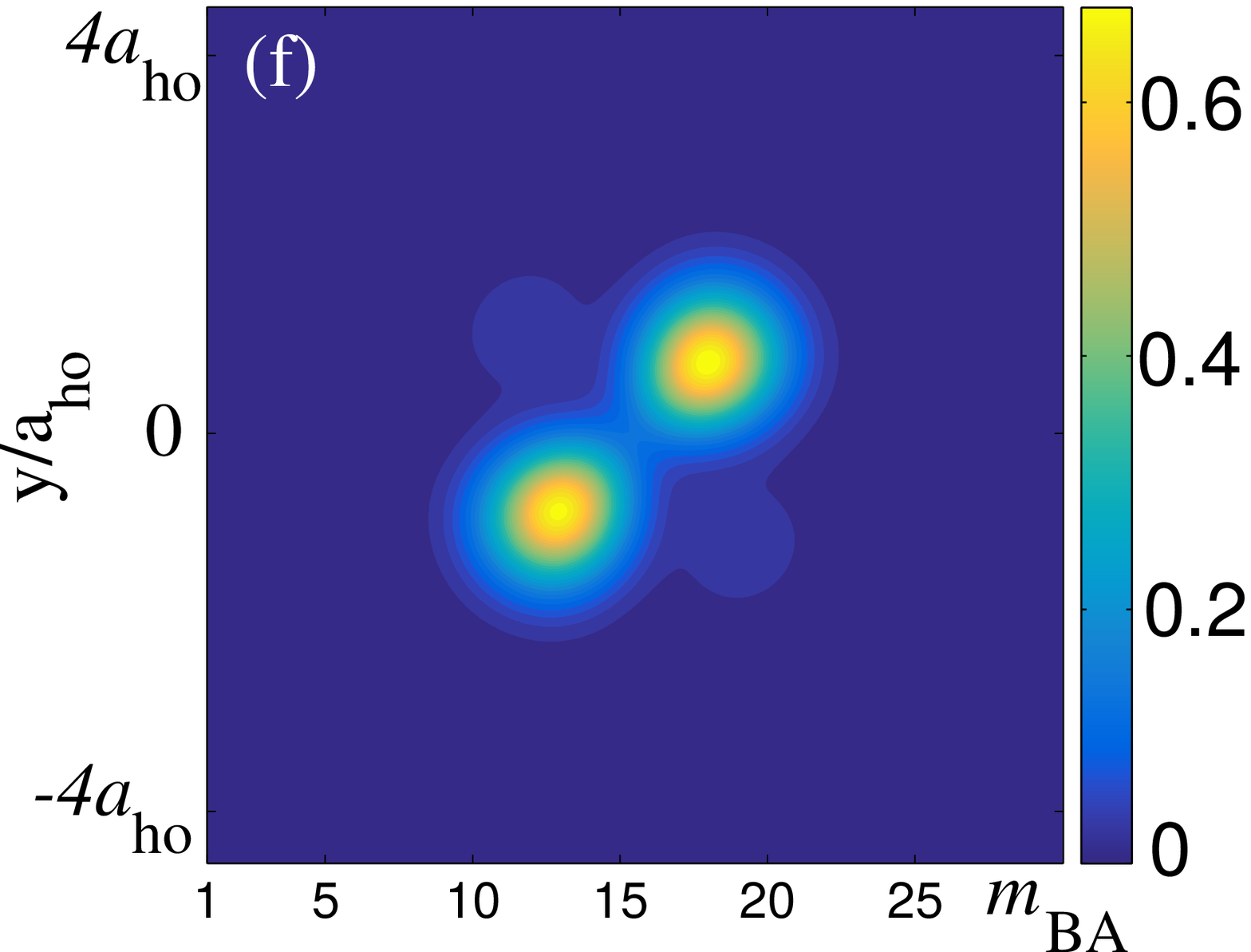}
\end{tabular}
\caption{One body density matrices for the bosons (left column) and the impurity (right column) for the ground state when $g=10$ and the mass ratio is close to the maximum of the entanglement entropy. Panels (a) and (d) correspond to $N_{\mathrm{A}}=2$ and $m_\mathrm{BA}=2.5$, panels (b) and (e) to $N_{\mathrm{A}}=3$ and $m_\mathrm{BA}=3.5$, panels (c) and (f) to $N_{\mathrm{A}}=4$ and $m_\mathrm{BA}=4.5$. \label{fig6}}
\end{figure}

From the density profiles one can observe that this localization of the B atom seems to occur abruptly as a function of $m_{\mathrm{BA}}$. This is more apparent from the calculation of the natural orbits occupations and the change in entanglement entropy as a function of $m_{\mathrm{BA}}$, which is shown in Fig.~\ref{fig5}. We plot these variables for different values of $N_{\mathrm{A}}$. 
First, the largest occupation of a natural orbital for 
both the bosons and the impurity is reduced as $m_{\mathrm{BA}}$ 
is increased as long as it is kept below a threshold value of the 
mass ratio
$m_{\mathrm{BA}}^{\mathrm{th}}$ [see Fig.~\ref{fig5} (a) and (b)]. We define the threshold mass ratio, 
$m_{\mathrm{BA}}^{\mathrm{th}}$, as the value of the mass ratio at which the largest natural occupation of a natural orbital for B reaches a minimum. This threshold mass depends 
on the number of bosons, $N_\textrm{A}$, and is increased for 
larger $N_{\mathrm{A}}$ for small values of $N_{\mathrm{A}}$ (see discussion on previous paragraph).  
For increasing $m_{\mathrm{BA}}$ while kept in the interval $[1,m_{\mathrm{BA}}^{\mbox{th}}]$ the entanglement entropy is increased [see Fig.~\ref{fig5} (c)]. Note that the entanglement entropy is calculated from the natural orbits occupations of B, so the maximum of the entanglement is related to the minimum of the largest natural orbital occupation for B.  Indeed, the overlap between the density profiles of the bosons and the impurity is yet large for $m_{\mathrm{BA}}>1$ but kept within this interval. In Fig.~\ref{fig6} we show the OBDMs for $N_{\mathrm{A}}=2,3$ and $4$ for a value of the mass ratio close to the maximum of the entanglement entropy [see Fig.~\ref{fig2}(c)]. These OBDMs show that the density profiles (the diagonals of the OBDMs) show a great overlap between the impurity and the A atomic cloud. In addition, the off-diagonal terms show that these are very correlated, as they correspond to the large value of the entanglement entropy.

For values of the mass ratio larger than $m_{\mathrm{BA}}^{\mbox{th}}$, the largest occupation of a natural orbital of both species grows toward a value close to 1. For the bosons this final value coincides with different values of $N_{\mathrm{A}}$. On the contrary, for B this value is smaller for larger values of $N_{\mathrm{A}}$. Further, the entanglement entropy drops quickly down. The final value of the entanglement entropy is larger for larger $N_{\mathrm{A}}$. Moreover, the density profile of the impurity shows that it localizes in the center of the trap, while the density profile of the bosons shows a minimum  where the impurity is localized. So as one increases the mass ratio above one, the first effect is to increase the correlations among the bosons and the impurity. The impurity has a tendency to occupy the center of the trap due to the increase in its mass, but the repulsion with the bosons keeps it towards the edges of the system. Remarkably, when the mass ratio is large enough as to produce the localization of the impurity in the center of the trap, the largest natural occupation of A tends to a value close to one. 
This in turn means that a single impurity is not able to fragment the bosons for large mass ratios. 
Indeed, the fact that the largest occupation of a natural 
orbital for B is smaller and that the entanglement entropy 
is larger for larger $N_{\mathrm{A}}$ shows that there is 
yet correlation between the bosons and the impurity which 
gets larger as $N_{\mathrm{A}}$ is increased.

\section{Conclusions}
\label{Sec:conc}
We have shown that when the impurity and up to four bosons interact with each other the correlation in the system grows as a function of interaction strength. Furthermore, we have shown that when the impurity is much  heavier than the bosons, then we have the least possible correlation. Moreover,  the entropy reaches a maximum when the mass ratio is somewhere between 3 and 4 for up to four bosonic particles and then falls off afterwards. Whether or not this holds for higher number of bosonic particles is an open question for future investigation. Here, we did not study the possibility of tuning the trapping frequency for the impurity, which is an experimentally feasible strategy, but we expect that it will also lead to a strongly correlated state for certain value of $\omega_{\mathrm{B}}$. The study on the effect of $\omega_{\mathrm{B}}$ or the combined effect of changing   $m_{\mathrm{BA}}$ together with $\omega_{\mathrm{B}}$ falls out of the scope of the paper and is left as an outlook. Also, a detailed study of the limit of $N_\textrm{A}\to\infty$ and how the  threshold mass changes as  $g\to\infty$ is left for future investigations. 

\section{Acknowledgments}
The authors thank Xiaoling Cui for reading and commenting on the manuscript. This work was funded by the Danish Council for Independent Research DFF Natural Sciences and the DFF Sapere Aude program.
M.A.G.-M. acknowledges support from EU grants OSYRIS (ERC-2013-AdG Grant No. 339106), SIQS (FP7-ICT-2011-9 No. 600645), EU STREP QUIC (H2020-FETPROACT-2014 No. 641122), EQuaM (FP7/2007-2013 Grant No. 323714), Spanish Ministry grant FOQUS (FIS2013-46768-P),  the Generalitat de Catalunya project 2014 SGR 874, the John Templeton Fundation, and Fundaci\'o Privada Cellex. Financial support from the Spanish Ministry of Economy and Competitiveness, through the "Severo Ochoa" Programme for Centres of Excellence in R\&D (SEV-2015-0522) is acknowledged.

\appendix
\section{Polaron Method}
\label{sec:app1}
In this section we explain briefly how we calculated the results for the 9 bosonic particles. The method is discussed in more details in \cite{dehkharghani2015b}, however, we also discuss some of the limitations in the method when it comes to the mass-imbalanced case.
The Hamiltonian is written as,
\begin{align}
H &= \frac{1}{m_{\mathrm{BA}}}\left(-\frac{1}{2}\frac{ d^2}{dy^2}\right) + m_\mathrm{BA} \frac{1}{2} y^2 \\
&+ \sum_{i=1}^{N_\mathrm{A}} \left[-\frac{1}{2}\frac{d^2}{dx_i^2} + \frac{1}{2} x_i^2\right] + g \sum_{i=1}^{N_\mathrm{A}} \delta(x_i-y),\nonumber
\end{align}
The $y$ and $x_i$ coordinates represent the impurity and the majority particles respectively as in Eq.~(\ref{eq:Hamiltonian}). Therefore we only remind that $m_{\mathrm{BA}}\equiv m_{\mathrm{B}}/m_{\mathrm{A}}$ and all energies are scaled by $\hbar\omega$ and all distances by the harmonic oscillator length $a_{\mathrm {ho}}=\sqrt{\hbar/m_{\mathrm{A}}\omega}$. We introduce an adiabatic decomposition of the total wave function of the form
\begin{align}
\Psi(y,x_1,\ldots,x_{N_\mathrm{A}})=\sum_j \phi_j(y)\Phi_j(x_1,\ldots,x_{N_\mathrm{A}}|y),
\end{align}
where $\Phi_j$ is a normalized eigenstate of the eigenproblem $\sum_{i=1}^{N_\mathrm{A}}h_0(x_i)\Phi_j=E_j(y)\Phi_j$, where $h_0(x_i)\equiv\left[-\frac{1}{2}\frac{d^2}{dx_i^2} + \frac{1}{2} x_i^2\right]$ and the eigenvalue problem depends parametrically on $y$. For a given $g$, we impose the condition that the total wave function must satisfy a delta-function boundary condition for $x_i=y$, $i=1,\ldots,N_\mathrm{A}$. This implies a discontinuity in the derivative of $\Phi_j$ whenever $x_i=y$ or $\Phi_j=0$ when $1/g \rightarrow 0$. Since we assume the bosonic particles are noninteracting with each other, we can write 
\begin{align}
\Phi_j(x_1,\ldots,x_{N_\mathrm{A}}|y)=\hat{S}\prod_{i=1}^{N_\mathrm{A}}f_{k_i}(x_i|y),
\end{align}
where $\hat{S}$ denotes the symmetrization operator and $f_{k_i}(x_i|y)$ is the $k$th normalized eigenstate of $h_0(x_i)$, which satisfies the delta-boundary condition. The index $j$ on $\Phi_j$ denotes the many different ways to distribute the $N_\mathrm{A}$ particles among the eigenstates of $h_0(x_i)$ with the appropriate boundary condition.
The Hamiltonian for $\phi_j(y)$, which has to be solved, can now be written
\begin{align*}
&\frac{-1}{m_{\mathrm{BA}}}\sum_j \left(Q_{ij}(y)\phi_j+P_{ij}(y)\frac{\partial \phi_j}{\partial y}\right) - \frac{1}{m_{\mathrm{BA}}} \frac{1}{2}\frac{ d^2}{dy^2} +\\& m_\mathrm{BA} \frac{1}{2} y^2\phi_i +E_i(y)\phi_i,
\end{align*}
where 
\begin{align}
&P_{ij}(y)=\langle \Phi_i|\frac{\partial}{\partial y}|\Phi_j\rangle_{x}&\\
&Q_{ij}(y)=\frac{1}{2}\langle\Phi_i|\frac{\partial^2}{\partial y^{2}}|\Phi_j\rangle_{x}.&
\end{align}
The subscript $x$ on the brackets denote integration over all $x_1,\ldots,x_{N_\mathrm{A}}$. Note that $P_{ii}=0$ and $Q_{ii}<0$ \cite{nielsen2001} and one can show that $P_{ij}$ and $Q_{ij}$ scales with $\sqrt{N_\mathrm{A}}$ while
$Q_{ii}$ scales with $N_\mathrm{A}$ \cite{dehkharghani2015a}. For large $N_\mathrm{A}$ we can thus neglect all but the $Q_{ii}$ terms. Particularly, if we only keep the first $Q_{11}$ then we have
\begin{align}
Q_{11}(y)=-\frac{1}{2}N_\mathrm{A}\left\langle\left(\frac{\partial f(x|y)}{\partial y}\right)^2\right\rangle_{x}.
\end{align}
Furthermore, $E_1(y)=N_\mathrm{A}\epsilon(y)$ by additivity.
Once we have determined the functions $f(x|y)$ and $\epsilon(y)$, see \cite{dehkharghani2015a, dehkharghani2015b}, we can compute the adiabatic potential for the ground state. The Schr{\"o}dinger equation for $\phi(y)$ is then
\begin{align}
&\frac{1}{m_{\mathrm{BA}}} \frac{N_\mathrm{A}}{2}\left\langle\left(\frac{\partial f(x|y)}{\partial y}\right)^2\right\rangle_x - \frac{1}{m_{\mathrm{BA}}} \frac{1}{2}\frac{ d^2}{dy^2} +m_\mathrm{BA} \frac{1}{2} y^2\phi_i + \nonumber \\& N_\mathrm{A}\epsilon(y)\phi_i = E \phi_i,
\label{polaroneffective}
\end{align}

The energy $E$ provides a variational upper bound to the exact energy. The energies computed via this method for the polaron are shown in Fig.~\ref{fig1} and
agree with the numerical exact diagonalization results in \cite{dehkharghani2015a} to within a few percent and is expected to agree better for larger number of particles.

However, the equation above, Eq.~(\ref{polaroneffective}), has some limitations in the strongly interacting case. The problem arises in the first term around $x=0$ as the impurity approaches zero from left and right. As the wave function $f(x|y)$ changes suddenly from right to left of zero this gives a huge contribution to the derivative resulting in a peak/wall in the middle of the effective potential for the impurity. The change in $f(x|y)$ and therefore the height of this wall is $g$ dependent. The stronger $g$, the bigger is the change in the $f(x|y)$ as the impurity is placed to the left and right side of $x=0$. But the first term in the equation is also $N_\mathrm{A}$ and $m_\mathrm{BA}$ dependent. As the impurity gets heavier the $1/m_\mathrm{BA}$ damps the huge contribution from the derivative, but the $N_\mathrm{A}$ on the other hand has the opposite effect like $g$. This means that when $g$ is strong, and we work with equal mass case, then the impurity is well separated and the physics of the impurity is not affected by this wall at all. But as one turns on for the mass imbalance, the impurity starts to move to the middle and starts to see this wall. All in all, the weak point of the method turns into a question of how big $g$ and $N_\mathrm{A}$ are relative to the mass ratio. On the other hand, if one knows that the impurity will be at the edges, then the method is very powerful to predict the energy and wave function for the polaron system. Otherwise an analysis of each contributing factors, $g$, $N_\mathrm{A}$ and $m_\mathrm{BA}$, has to be made in order to validate the method. For instance, in the $N_\mathrm{A}=4$ and $g<2$ case, the method works perfectly for any $m_\mathrm{BA}$.


\begin{thebibliography}{99}

\bibitem{paredes2004} B. Paredes {\it et al.}, Nature {\bf 429}, 277-281 (2004).
\bibitem{kino2004} T. Kinoshita, T. Wenger, and D.~S. Weiss, Science {\bf 305}, 1125-1128 (2004).
\bibitem{kinoshita2006} T. Kinoshita, T. Wenger, and D.~S. Weiss, Nature {\bf 440}, 900-903 (2006).
\bibitem{haller2009} E. Haller {\it et al.}, Science {\bf 325}, 1224-1227 (2009).
\bibitem{serwane2011} F. Serwane {\it et al.}, Science {\bf 332}, 336-338 (2011).
\bibitem{gerhard2012} G. Z{\"u}rn {\it et al.}, Phys. Rev. Lett. {\bf 108}, 075303 (2012).
\bibitem{wenz2013} A.~N. Wenz {\it et al.}, Science {\bf 342}, 457 (2013).

\bibitem{recati2003} A. Recati, P.~O. Fedichev, W. Zwerger, and P. Zoller, Phys. Rev. Lett. {\bf 90}, 020401 (2003).

\bibitem{eisenberg2002} E. Eisenberg and E.~H. Lieb, Phys. Rev. Lett. {\bf 89}, 220403 (2002).
\bibitem{nachtergaele2005} B. Nachtergaele and S. Shannon, Phys. Rev. Lett. {\bf 94}, 057206 (2005).
\bibitem{deuretzbacher2008} F. Deuretzbacher {\it et al.}, Phys. Rev. Lett. {\bf 100} 160405 (2008)
\bibitem{massignan2015} P. Massignan, J. Levinsen, and M.~M. Parish, Phys. Rev. Lett. {\bf 115}, 247202 (2015).
\bibitem{yang2015} L. Yang and X. Cui, Phys. Rev. A {\bf 93}, 013617 (2016).

\bibitem{landau1933} L.~D. Landau, Phys. Z. Sowjetunion {\bf 3}, 644 (1933).
\bibitem{pekar1948} L.~D. Landan and S.~I. Pekar, J. Exp. Theor. Phys. {\bf 18}, 419 (1948).
\bibitem{kondo1964} J. Kondo, Prog. Theor. Phys. {\bf 32}, 37 (1964).

\bibitem{cucchietti2006} F.~M.~Cucchietti and E.~Timmermans, Phys. Rev. Lett. {\bf 94}, 210401 (2006).

\bibitem{kalas2006} R.~M. Kalas and D. Blume, Phys. Rev. A {\bf 73}, 043608 (2006).
\bibitem{bruderer2008} M. Bruderer, W. Bao, and D. Jaksch, Europhys. Lett. {\bf 82}, 30004 (2008).

\bibitem{Tempere2013} J.~Tempere, W.~Casteels, M.~K.~Oberthaler, S.~Knoop, E.~Timmermans and J.~T.~Devrees, Phys. Rev. B {\bf 80}, 184504 (2009).

\bibitem{rath2013} S.~P. Rath and R. Schmidt, Phys. Rev. A {\bf 88}, 053632 (2013).
\bibitem{li2014} W. Li and S. Das Sarma, Phys. Rev. A {\bf 90}, 013618 (2014).
\bibitem{grusdt2015} F. Grusdt and E. Demler, arXiv:1510.04934 (2015).
\bibitem{mehta2014} N.~P. Mehta, Phys. Rev. A {\bf 89}, 052706 (2014).
\bibitem{artem2015b} A.~G. Volosniev, H.-W. Hammer, and N.~T. Zinner, Phys. Rev. A {\bf 92}, 023623 (2015).
\bibitem{mehta2015} N.~P. Mehta, Phys. Rev. A {\bf 92}, 043616 (2015).


\bibitem{dehkharghani2015a} A. S. Dehkharghani {\it et al.}, Scientific Reports {\bf 5}, 10675 (2015).
\bibitem{dehkharghani2015b} A. S. Dehkharghani, A. G. Volosniev, and N. T. Zinner, Phys. Rev. A {\bf 92}, 031601(R) (2015).

\bibitem{Zollner:08a}
S. Z\"{o}llner, H.-D. Meyer, and P. Schmelcher,
Phys.~Rev.~A \textbf{78}, 013629 (2008).

\bibitem{Garcia-March:14} M.~A. Garcia-March {\it et al.}, New J. Phys. {\bf 16} (2014) 103004.

\bibitem{Garcia-March:14b}
M.~A. Garcia-March {\it et al.}
Phys. Rev. A \textbf{90}, 063605 (2014).

\bibitem{Zinner:13}
N. T. Zinner {\it et al.}, Europhysics Letters {\bf 107} 60003 (2014).

\bibitem{harshman2012} N.~L.~Harshman, Phys. Rev. A {\bf 86}, 052122 (2012).
\bibitem{Harshman:15} N.~L. Harshman, Few-Body Syst. {\bf 57}, 11 (2016). N.~L. Harshman, Few-Body Syst. {\bf 57}, 45 (2016).

\bibitem{yurovsky2014} V.~A.~Yurovsky, Phys. Rev. Lett. {\bf 113}, 200406 (2014).

\bibitem{Garcia-march:12}
M.~A. Garcia-March and Th. Busch, Phys. Rev. A \textbf{87}, 063633 (2013).
\bibitem{Garcia-March:13} M.~A. Garcia-March {\it et al.}, Phys. Rev. A \textbf{88}, 063604 (2013).


\bibitem{pecak2015} D. Pecak, M. Gajda, and T. Sowi{\'n}ski, New J. Phys. {\bf 18}, 013030 (2016).

\bibitem{loft2015} N.~J.~S.~Loft {\it et al.}, EPJD {\bf 69:}65 (2015).

\bibitem{Giraud2009}  S. Giraud and R. Combescot, Phys. Rev. A {\bf 79}, 043615 (2009).

\bibitem{nielsen2001} E. Nielsen, D.~V. Fedorov, A.~S. Jensen, and E. Garrido, Phys. Rep. {\bf 347}, 373-459 (2001).



\end{thebibliography}
\end{document}